\documentclass[floatfix,twocolumn,showpacs,preprintnumbers,amsmath,amssymb]{revtex4}
\usepackage{dsfont}
\usepackage{cases}
\usepackage{graphicx}
\usepackage{dcolumn}
\usepackage{bm}
\usepackage{amsmath}
\usepackage{subfigure}

\begin{document}

\title{ Non-Markovian dynamics and quantum interference of open three-level quantum systems}
\author{Hao-Sheng Zeng}%
\email{hszeng@hunnu.edu.cn}
\author{Yu-Kun Ren}
\author{Zhi He}
\affiliation{Key Laboratory of Low-Dimensional Quantum Structures and Quantum
Control of Ministry of Education, Synergetic Innovation Center for Quantum Effects and Applications, and Department of Physics,
Hunan Normal University, Changsha 410081, China}%

\date{\today}

\begin{abstract}
The exactly analytical solutions for the dynamics of the dissipative three-level V-type and $\Lambda$-type atomic systems in the vacuum Lorentzian environments are presented. Quantum interference phenomenon between the two transitions in V-type atomic system is studied and the intuitive interference conditions are derived. For the dissipative $\Lambda$-type atomic system, we demonstrate that the similar quantum interference phenomenon does not exist. Finally, we study the dynamical evolution of quantum Fisher information, quantum entanglement and quantum coherence for the V-type atomic system. We find that quantum interference plays a positive role to the protection of quantum entanglement and quantum coherence. The difference between the two typical measures of the quantum coherence is also demonstrated in the considered systems.
\end{abstract}
\maketitle

\section{Introduction}
The study of open quantum systems is very
important, because no realistic quantum system is completely isolated from its surroundings. It is not only relevant for better understanding of
quantum theory, but also fundamental for various modern applications
of quantum mechanics, especially for quantum communication,
cryptography and computation \cite{Nielsen}. The early study of
dynamics of open quantum systems usually involves the application
of the Born-Markov approximation, that is, neglects all
the memory effects, leading to a master equation which can be cast
in the so-called Lindblad form \cite{Lindblad,Gorini}. Master
equation in Lindblad form can be characterized by the fact that the
dynamics of the system satisfies both the semigroup property and the
complete positivity, thus ensuring the preservation of positivity of
the density matrix during the time evolution. We usually attribute
this kind of dynamical process to the
well-known Markovian one.

However, people found that Many relevant physical systems,
such as the quantum optical system \cite{Breuer3} and the nanoscale solid-state quantum system\cite{Buluta2011,You2011}, could not be described simply by the Markovian dynamics. Similarly, quantum chemistry
\cite{Shao} and the excitation transfer of a biological system
\cite{Chin} also need to be treated as non-Markovian processes.
Quantum non-Markovian dynamics can lead to some interesting phenomena such as quantum correlation and coherence trapping \cite{Bellomo,Haikka2013,Addis}, correlation quantum beat \cite{Zeng}, and has extensively possible applications in quantum metrology \cite{Chin1}, quantum communication \cite{Vasile,Laine1,Bylicka,Tangning1}, quantum control \cite{Schmidt}.
Because of these distinctive
properties and extensive applications, more and more attention and
interest have been devoted to the study of non-Markovian process
of open systems, including the measure of non-Markovianity
\cite{Breuer,Laine,Rivas,Wolf,Lu,Luo20121,Chru2014,Hall2014,Chru2017,Song2015,Chen2016,Dhar2015,Paula2016,Hezhi2017}, the positivity
\cite{Breuer1,Shabani,Breuer2}, and some other dynamical properties
\cite{Haikka,Chang,Krovi,Chru,Haikka1,Wibmann2015,Bae2016,Bylicka2017,Zeng2015} and approaches
\cite{Jing,Koch,Wu} of non-Markovian processes. Experimentally, the
simulation \cite{Xu1,Xu2,Biheng,Jianshun,Fanchini2014,Bernardes2015} of non-Markovian environments has been
realized.

The dynamics of open quantum systems is very sophisticated, and only very rare of which can be solved exactly.  The case of a two-level atom dissipating in a vacuum environment is one of the few examples that can be solved exactly. Duo to the advantage of the exactly analytical solution, the dissipative two-level system becomes the paradigm for the investigation of non-Markovian dynamics hotted up very recently. Multilevel open quantum systems, especially multilevel dissipative systems, due to their complexity,  are relatively seldom involved. Though several works have already involved the study of dissipative three-level systems \cite{Dalton2001,Gaoxiang2012}, the concrete expression of the exactly analytical solution that relates the evolved state to its initial state (or Kraus representation) has not yet been established explicitly. As the significance of theoretical researches such as quantum interference\cite{Scully1997}, and the potential applications such as the quantum cryptography \cite{Brub,Cerf} and the fault-tolerant quantum computation and quantum error correction \cite{Knill}, it is very worthwhile to pour more efforts into the study of dynamics of  multilevel open quantum systems. Motivated by these facts, we will in this paper present the exactly analytical solutions for the dissipative three-level V-type and $\Lambda$-type atoms in the vacuum Lorentzian environments, and make some researches on the dynamical properties such as the quantum interference, the evolutions of quantum Fisher information, quantum entanglement and quantum coherence.

The paper is organized as follows. In Sec. II, we introduce respectively the models for the V-type and $\Lambda$-type atoms interacting with vacuum environments and present the exactly analytical solutions. In Sec.III, we study the phenomenon of quantum interference between different decaying channels for the two kinds of atomic models. In Secs. IV and V, we study respectively the evolution of quantum Fisher information, quantum entanglement and quantum coherence, for the dissipative V-type atomic system, especially highlighting the roles of quantum interference. Finally, we give the conclusions in Sec. VI.

\section{Dynamical models and their solutions}

\subsection{V-type three-level atom}
Consider a V-type three-level atom with transition frequencies $\omega_{1}$ and $\omega_{2}$ (see Fig.1a), which
is embedded in a zero-temperature bosonic reservoir modeled by an
infinite chain of quantum harmonic oscillators. The Hamiltonian for the whole system may be written as
\begin{eqnarray}
  H_{V}&=&\omega_{1}|1\rangle\langle1|+\omega_{2}|2\rangle\langle2|+\sum_{k}\omega_{k}b^{\dagger}_{k}b_{k}\\
 \nonumber &+&\sum_{k}\left[g_{1k}b_{k}|1\rangle\langle0|+g_{2k}b_{k}|2\rangle\langle0|+{\rm h.c.}\right].
\end{eqnarray}
Where $\omega_{1}$ and $\omega_{2}$ are respectively the energy ($\hbar=1$) of levels $|1\rangle$ and $|2\rangle$, and we set the energy of level $|0\rangle$ to be zero. $b_{k}$ and $b_{k}^{\dagger}$ are the annihilation and creation operators for
the $k$-th harmonic oscillator of the reservoir, $g_{1k}$ and $g_{2k}$ are the coupling strengthes between reservoir and the two transition channels respectively.

\begin{figure}
\vspace{-1.5cm}
\hspace{-1.0cm}
\includegraphics[width=4.0in,height=16cm]{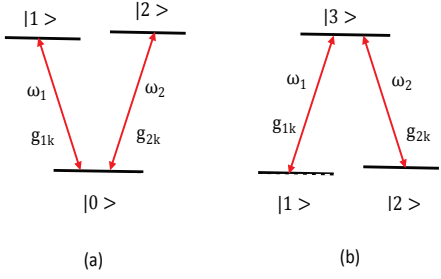}
\vspace{-10cm}
\caption{ Schematic diagram of energy level for (a) V-type atom and (b) $\Lambda$-type atom. }
\end{figure}

Suppose that the initial state of the atom plus its environment is
\begin{equation}
    |\Psi(0)\rangle=\left[c_{0}(0)|0\rangle+c_{1}(0)|1\rangle+c_{2}(0)|2\rangle\right]\otimes|0\rangle_{R},
\end{equation}
where $|0\rangle_{R}$ denotes the vacuum state of environment. Employing the conservativeness of excitation numbers of Jaynes-Cummings model, the dynamical state at any time $t$ may be written as
\begin{eqnarray}
   \nonumber |\Psi(t)\rangle &=& \left[c_{0}(t)|0\rangle+c_{1}(t)|1\rangle+c_{2}(t)|2\rangle\right]\otimes|0\rangle_{R}\\
    &+&\sum_{k}c_{k}(t)|0\rangle\otimes|1_{k}\rangle_{R},
\end{eqnarray}
where $|1_{k}\rangle_{R}$ indicates that there is a photon in the $k$th mode of the environment.
Tracing over the environmental degrees of freedom, then the reduced state of the atom in its natural bases is
\begin{small}
\begin{equation}
    \rho_{s}(t)=\left[
                        \begin{array}{ccc}
                          1-|c_{1}(t)|^{2}-|c_{2}(t)|^{2} & c_{0}(t)c_{1}^{*}(t) & c_{0}(t)c_{2}^{*}(t) \\
                          c_{0}^{*}(t)c_{1}(t) & |c_{1}(t)|^{2} & c_{1}(t)c_{2}^{*}(t) \\
                          c_{0}^{*}(t)c_{2}(t) & c_{1}^{*}(t)c_{2}(t) & |c_{2}(t)|^{2} \\
                        \end{array}
                      \right]
\end{equation}
\end{small}

The evolution of coefficients $c_{i}(t)$ is determined by the Schr\"{o}dinger equation ${\rm i}\partial|\Psi(t)\rangle/\partial t=H_{V}|\Psi(t)\rangle$, which satisfy the following set of equations
\begin{subnumcases}{}
  c_{0}(t) = c_{0}(0),\\
  {\rm i}\dot{c}_{1}(t) = \omega_{1}c_{1}(t)+\sum_{k}g_{1k}c_{k}(t),\\
   {\rm i}\dot{c}_{2}(t) = \omega_{2}c_{2}(t)+\sum_{k}g_{2k}c_{k}(t),\\
  {\rm i}\dot{c}_{k}(t) = \omega_{k}c_{k}(t)+g_{1k}^{*}c_{1}(t)+ g_{2k}^{*}c_{2}(t).
\end{subnumcases}
Formally integrating equation (5d) and plugging it into (5b)-(5c), one obtains
\begin{eqnarray}
 \nonumber \dot{c}_{1}(t)&=& -{\rm i}\omega_{1}c_{1}(t)-\sum_{k}|g_{1k}|^{2}\int_{0}^{t} {\rm d}\tau c_{1}(\tau)e^{-{\rm i}\omega_{k}(t-\tau)} \\
  &-& \sum_{k}g_{1k}g_{2k}^{*}\int_{0}^{t} {\rm d}\tau c_{2}(\tau)e^{-{\rm i}\omega_{k}(t-\tau)},\\
  \nonumber \dot{c}_{2}(t) &=& -{\rm i}\omega_{2}c_{2}(t)-\sum_{k}|g_{2k}|^{2}\int_{0}^{t} {\rm d}\tau c_{2}(\tau)e^{-{\rm i}\omega_{k}(t-\tau)} \\
  &-& \sum_{k}g_{2k}g_{1k}^{*}\int_{0}^{t} {\rm d}\tau c_{1}(\tau)e^{-{\rm i}\omega_{k}(t-\tau)}.
\end{eqnarray}
In the above deduction, we have used the initial condition $c_{k}(0)=0$. Without loss of generality, we assume in the following $g_{1k}$ and $g_{2k}$ to be real. In the continuum limitation of environmental modes $\sum_{k}g_{ik}g_{jk}\rightarrow \int {\rm d}\omega J_{ij}(\omega)$, where $J_{ij}(\omega)$ with $i,j=1,2$ is the spectral density function, one obtains
 \begin{eqnarray}
  \nonumber \dot{c}_{1}(t)&=&-{\rm i}\omega_{1}c_{1}(t)-\int_{0}^{t}{\rm d}\tau f_{11}(t-\tau)c_{1}(\tau)\\
   &-&\int_{0}^{t}{\rm d}\tau f_{12}(t-\tau)c_{2}(\tau),\\
  \nonumber \dot{c}_{2}(t)&=&-{\rm i}\omega_{2}c_{2}(t)-\int_{0}^{t}{\rm d}\tau f_{22}(t-\tau)c_{2}(\tau)\\
   &-&\int_{0}^{t}{\rm d}\tau f_{21}(t-\tau)c_{1}(\tau),
\end{eqnarray}
where $f_{ij}(t-\tau)=\int{\rm d}\omega J_{ij}(\omega)e^{-{\rm i}\omega(t-\tau)}$ is the two-point correlation function of environment.
Denoting $c_{i}(p)$ and $f_{ij}(p)$ the Laplace transform of $c_{i}(t)$ and $f_{ij}(t)$ respectively, then equations (8) and (9) lead to
\begin{eqnarray}
  c_{1}(p)=\frac{[p+{\rm i}\omega_{2}+f_{22}(p)]c_{1}(0)-f_{12}(p)c_{2}(0)}{Q(p)},\\
  c_{2}(p)=\frac{[p+{\rm i}\omega_{1}+f_{11}(p)]c_{2}(0)-f_{21}(p)c_{1}(0)}{Q(p)},
\end{eqnarray}
with $Q(p)=[p+{\rm i}\omega_{1}+f_{11}(p)][p+{\rm i}\omega_{2}+f_{22}(p)]-f_{12}(p)f_{21}(p)$.
In principle, the inverse Laplace transform of these equations gives the time evolution of $c_{1}(t)$ and $c_{2}(t)$.

To go further, we need to specify the spectral density of the environment. As an exemplification, we choose the Lorentzian spectrum
\begin{equation}
    J_{ij}(\omega)=\frac{1}{2\pi}\cdot\frac{\gamma_{ij}\lambda^{2}}{(\omega_{0}-\omega)^{2}+\lambda^{2}},
\end{equation}
where $\omega_{0}$ is the central frequency and $\lambda$ defines the spectral width. The parameter $\gamma_{ii}\equiv \gamma_{i}$ with $i=1,2$ describe the spontaneous emission rates of level $|i\rangle$, and $\gamma_{ij}$ with $i\neq j$ describe the correlation between the dipole moments
corresponding to the two transition channels in Fig.1(a). When the dipole moments of the two transitions are parallel, the relation $\gamma_{12}=\gamma_{21}=\sqrt{\gamma_{1}\gamma_{2}}$ is met. In this paper, we consider only this case.

For the Lorentzian spectrum, the correlation function may be calculated as $f_{ij}(t-\tau)=\frac{\gamma_{ij}\lambda}{2}e^{-M(t-\tau)}$ and the corresponding Laplace transform reads $f_{ij}(p)=B_{ij}/(p+M)$ with $B_{ij}=\gamma_{ij}\lambda/2$ and $M=\lambda+{\rm i}\omega_{0}$. If the dipole moments of the two transitions are parallel, one also has
$B_{11}B_{22}-B_{12}B_{21}=0$. Taking these results into consideration, equations (10)-(11) become as,
\begin{eqnarray}
  c_{1}(p)=\frac{[(p+{\rm i}\omega_{2})(p+M)+B_{22}]c_{1}(0)-B_{12}c_{2}(0)}{p^{3}+h_{1}p^{2}+h_{2}p+h_{3}},\\
  c_{2}(p)=\frac{[(p+{\rm i}\omega_{1})(p+M)+B_{11}]c_{2}(0)-B_{21}c_{1}(0)}{p^{3}+h_{1}p^{2}+h_{2}p+h_{3}},
\end{eqnarray}
where $h_{1}=M+{\rm i}(\omega_{1}+\omega_{2})$, $h_{2}=B_{11}+B_{22}-\omega_{1}\omega_{2}+{\rm i}M(\omega_{1}+\omega_{2})$, $h_{3}=-\omega_{1}\omega_{2}M+{\rm i}(\omega_{1}B_{22}+\omega_{2}B_{11})$.

Observing that the numerator and denominator of Eqs. (13)-(14) are the second and third order polynomials of $p$, if the roots $b_{i}$ with ($i=1,2,3$) of the polynomial equation $p^{3}+h_{1}p^{2}+h_{2}p+h_{3}=0$ are non-degenerate, then we can make the decomposition
\begin{subnumcases}{}
  c_{1}(p)=\sum_{i=1}^{3}\frac{D_{i}}{p-b_{i}},\\
  c_{2}(p)=\sum_{i=1}^{3}\frac{D'_{i}}{p-b_{i}}.
\end{subnumcases}
For the degenerate case, the decomposition has different form which is discussed in Appendix A. Noting that $b_{i}$ are determined by the atomic and environmental structure parameters, the degenerative probability is very small and may be avoided by adjusting these structure parameters. The parameters $D_{i}$ ($D'_{i}$) are the residues of $c_{1}(p)$ ($c_{2}(p)$) at the points $p=b_{i}$, which in our problem may be written as
\begin{subnumcases}{}
  D_{i}=E_{i}c_{1}(0)+F_{i}c_{2}(0),\\
  D'_{i}=G_{i}c_{2}(0)+H_{i}c_{1}(0),
\end{subnumcases}
with
\begin{eqnarray}
 \nonumber E_{i}&=&\frac{(b_{i}+{\rm i}\omega_{2})(b_{i}+M)+B_{22}}{3b_{i}^{2}+2h_{1}b_{i}+h_{2}},\\
 \nonumber F_{i}&=&-\frac{B_{12}}{3b_{i}^{2}+2h_{1}b_{i}+h_{2}},\\
 \nonumber G_{i}&=&\frac{(b_{i}+{\rm i}\omega_{1})(b_{i}+M)+B_{11}}{3b_{i}^{2}+2h_{1}b_{i}+h_{2}},\\
 \nonumber H_{i}&=&-\frac{B_{21}}{3b_{i}^{2}+2h_{1}b_{i}+h_{2}}.
\end{eqnarray}
Finally the inverse Laplace transform of Eqs. (15a)-(15b) gives
\begin{subnumcases}{}
  c_{1}(t)=E(t)c_{1}(0)+F(t)c_{2}(0),\\
  c_{2}(t)=G(t)c_{2}(0)+H(t)c_{1}(0),
\end{subnumcases}
with
\begin{eqnarray}
 \nonumber E(t)&=&\sum_{i=1}^{3}E_{i}e^{b_{i}t},\\
 \nonumber F(t)&=&\sum_{i=1}^{3}F_{i}e^{b_{i}t},\\
 \nonumber G(t)&=&\sum_{i=1}^{3}G_{i}e^{b_{i}t},\\
 \nonumber H(t)&=&\sum_{i=1}^{3}H_{i}e^{b_{i}t}.
\end{eqnarray}
So far, we have completed the solving process of the open V-type three-level atom and the exact analytical results are given by equations (4),(5a),(17a)-(17b).

\subsection{$\Lambda$-type three-level atom}
Besides open V-type three-level atoms, the dynamics of an open $\Lambda$-type three-level atom dissipating in a zero-temperature bosonic reservoir can also be solved exactly.
Fig.1b is the energy level structure of a $\Lambda$-type three-level atom with $|1\rangle$ and $|2\rangle$ the ground and meta-stable states, and $|3\rangle$ the excited state. The Hamiltonian for the atom plus its environment can be written as
\begin{eqnarray}
  H_{\Lambda}&=&-\omega_{1}|1\rangle\langle1|-\omega_{2}|2\rangle\langle2|+\sum_{k}\omega_{k}b^{\dagger}_{k}b_{k}\\
 \nonumber &+&\sum_{k}\left[g_{1k}b_{k}|3\rangle\langle1|+g_{2k}b_{k}|3\rangle\langle2|+{\rm h.c.}\right].
\end{eqnarray}
Here $\omega_{1}$ and $\omega_{2}$ are respectively the frequencies of the transitions $|1\rangle\leftrightarrow |3\rangle$ and $|2\rangle\leftrightarrow |3\rangle$, and we set the energy of level $|3\rangle$ to be zero. The meanings of other symbols are
the same as before.

Suppose that the initial state of the atom plus its environment is
\begin{equation}
    |\Upsilon(0)\rangle=\left[c_{1}(0)|1\rangle+c_{2}(0)|2\rangle+c_{3}(0)|3\rangle\right]\otimes|0\rangle_{R},
\end{equation}
with $|0\rangle_{R}$ the vacuum state of environment, then the dynamical state at any time $t$ may be written as
\begin{eqnarray}
   |\Upsilon(t)\rangle &=& \left[c_{1}(t)|1\rangle+c_{2}(t)|2\rangle+c_{3}(t)|3\rangle\right]\otimes|0\rangle_{R}\\
   \nonumber &+&\sum_{k}c_{k}(t)|1\rangle\otimes|1_{k}\rangle_{R}+\sum_{k}d_{k}(t)|2\rangle\otimes|1_{k}\rangle_{R}.
\end{eqnarray}
As before $|1_{k}\rangle_{R}$ indicates the single-photon state of the $k$th-mode environment.
The reduced state of the atom in its natural bases now becomes
\begin{widetext}
\begin{equation}
    \varrho_{s}(t)=\left[
                        \begin{array}{ccc}
                          |c_{1}(t)|^{2}+\sum_{k}|c_{k}(t)|^{2} & c_{1}(t)c_{2}^{*}(t)+\sum_{k}c_{k}(t)d_{k}^{*}(t) & c_{1}(t)c_{3}^{*}(t) \\
                          c_{2}(t)c_{1}^{*}(t)+\sum_{k}d_{k}(t)c_{k}^{*}(t) & |c_{2}(t)|^{2}+\sum_{k}|d_{k}(t)|^{2} & c_{2}(t)c_{3}^{*}(t) \\
                          c_{3}(t)c_{1}^{*}(t) & c_{3}(t)c_{2}^{*}(t) & |c_{3}(t)|^{2} \\
                        \end{array}
                      \right].
\end{equation}
\end{widetext}
Here the evolution of coefficients is determined by following set of equations
\begin{subnumcases}{}
  {\rm i}\dot{c}_{1}(t) =-\omega_{1} c_{1}(t),\\
  {\rm i}\dot{c}_{2}(t) = -\omega_{2}c_{2}(t),\\
   {\rm i}\dot{c}_{3}(t) = \sum_{k}g_{1k}c_{k}(t)+\sum_{k}g_{2k}d_{k}(t),\\
  {\rm i}\dot{c}_{k}(t) = -\omega_{1}c_{k}(t)+\omega_{k}c_{k}(t)+g_{1k}^{*}c_{3}(t),\\
  {\rm i}\dot{d}_{k}(t) = -\omega_{2}d_{k}(t)+\omega_{k}d_{k}(t)+g_{2k}^{*}c_{3}(t).
\end{subnumcases}
Obviously, the solving process of this problem is more complex than the case of V-type atom, because we need to know the evolution of $c_{k}(t)$ and $d_{k}(t)$, besides $c_{1}(t)$, $c_{2}(t)$ and $c_{3}(t)$. The evolution of $c_{1}(t)$ and $c_{2}(t)$ can be easily obtained
\begin{subnumcases}{}
  c_{1}(t) = c_{1}(0)e^{{\rm i}\omega_{1}t},\\
  c_{2}(t) = c_{2}(0)e^{{\rm i}\omega_{2}t}.
\end{subnumcases}
To get the evolution of the other coefficients, we formally integrate eqs.(22d)-(22e) in the condition $c_{k}(0)=d_{k}(0)=0$ and get
\begin{subnumcases}{}
  c_{k}(t) =-{\rm i}\int_{0}^{t}{\rm d}\tau g_{1k}^{*}c_{3}(\tau)e^{{\rm i}(\omega_{1}-\omega_{k})(t-\tau)},\\
  d_{k}(t) =-{\rm i}\int_{0}^{t}{\rm d}\tau g_{2k}^{*}c_{3}(\tau)e^{{\rm i}(\omega_{2}-\omega_{k})(t-\tau)}.
\end{subnumcases}
By plugging them into Eq. (22c), as well as using the continuum limitation $\sum_{k}g_{ik}g_{jk}\rightarrow \int {\rm d}\omega J_{ij}(\omega)$ with $i,j=1,2$,  we obtain
 \begin{equation}
  \dot{c}_{3}(t)=-\int_{0}^{t}{\rm d}\tau [f_{1}(t-\tau)+f_{2}(t-\tau)]c_{3}(\tau),
\end{equation}
here the correlation function is defined by $f_{j}(t-\tau)=\int{\rm d}\omega J_{jj}(\omega)e^{{\rm i}(\omega_{j}-\omega)(t-\tau)}$.

Now we also assume the Lorentzian spectrum Eq.(12) with $\gamma_{jj}\equiv \gamma_{j}$ ($j=1,2)$ that describe the spontaneous emissions of levels $|3\rangle$ to $|j\rangle$. Assume that the transitions have parallel dipole moments so that $\gamma_{12}=\gamma_{21}=\sqrt{\gamma_{1}\gamma_{2}}$. Under these conditions, we have $f_{j}(t-\tau)=\frac{\gamma_{j}\lambda}{2}e^{-M_{j}(t-\tau)}$
with $M_{j}=\lambda+{\rm i}(\omega_{0}-\omega_{j})$. The solution of Eq.(25) may be written as
\begin{equation}
  c_{3}(t)=\xi(t)c_{3}(0),
\end{equation}
with $\xi(t)=(\mathfrak{D}_{1}e^{\mathfrak{b}_{1}t}+\mathfrak{D}_{2}e^{\mathfrak{b}_{2}t}+\mathfrak{D}_{3}e^{\mathfrak{b}_{3}t})$. Here $\mathfrak{b}_{i}$ are the roots of equation $R(p)=p^{3}+(M_{1}+M_{2})p^{2}+(M_{1}M_{2}+\frac{\gamma_{1}\lambda}{2}+\frac{\gamma_{2}\lambda}{2})p+\frac{\gamma_{1}\lambda}{2}M_{2}+\frac{\gamma_{2}\lambda}{2}M_{1}=0$, which are assumed non-degenerate. As the degenerative probability is very small and can always be avoided by adjusting the structure parameters,
we thus have no longer presented the expresses in detail. The coefficients $\mathfrak{D}_{i}$ are given by
\begin{equation}
  \mathfrak{D}_{i}=\frac{(\mathfrak{b}_{i}+M_{1})(\mathfrak{b}_{i}+M_{2})}{3\mathfrak{b}_{i}^{2}+2(M_{1}+M_{2})\mathfrak{b}_{i}+M_{1}M_{2}+\frac{\gamma_{1}\lambda}{2}+\frac{\gamma_{2}\lambda}{2}}.
\end{equation}

Having $c_{3}(t)$ in hand, we can then obtain the evolution of $c_{k}(t)$ and $c_{k}(t)$ in terms of Eqs.(24a)-(24b). In the continuum limitation $\sum_{k}g_{ik}g_{jk}\rightarrow \int {\rm d}\omega J_{ij}(\omega)$ with $i,j=1,2$, we have
\begin{small}
\begin{eqnarray}
  \sum_{k}|c_{k}(t)|^{2}&=&\int_{0}^{t}{\rm d}\tau\int_{0}^{t}{\rm d}\tau' f_{1}(\tau'-\tau)c_{3}(\tau)c_{3}^{*}(\tau'),\\
  \sum_{k}|d_{k}(t)|^{2}&=&\int_{0}^{t}{\rm d}\tau\int_{0}^{t}{\rm d}\tau' f_{2}(\tau'-\tau)c_{3}(\tau)c_{3}^{*}(\tau'),
\end{eqnarray}
\end{small}
and
\begin{widetext}
\begin{eqnarray}
 \nonumber \sum_{k}c_{k}(t)d_{k}^{*}(t) &=& \int_{0}^{t}{\rm d}\tau\int_{0}^{t}{\rm d}\tau'\int{\rm d}\omega J_{12}(\omega)e^{{\rm i}\omega(\tau-\tau')}e^{{\rm i}(\omega_{1}t-\omega_{2}t-\omega_{1}\tau+\omega_{2}\tau')}c_{3}(\tau)c_{3}^{*}(\tau') \\
  &=& \frac{\gamma_{12}\lambda}{2}e^{{\rm i}(\omega_{1}-\omega_{2})t}\int_{0}^{t}{\rm d}\tau \int_{0}^{t}{\rm d}\tau' c_{3}(\tau)c_{3}^{*}(\tau')e^{-\lambda|\tau'-\tau|+{\rm i}(\delta_{1}\tau-\delta_{2}\tau')},
\end{eqnarray}
\end{widetext}
Where $f_{j}(\tau'-\tau)=\int{\rm d}\omega J_{jj}(\omega)\exp{[{\rm i}(\omega_{j}-\omega)(\tau'-\tau)]}=\frac{\gamma_{j}\lambda}{2}\exp{[-\lambda|\tau'-\tau|-{\rm i}\delta_{j}(\tau'-\tau)]}$, $\delta_{j}=\omega_{0}-\omega_{j}$,
and we have used $$\int {\rm d}\omega J_{12}(\omega)e^{{\rm i}\omega(\tau-\tau')}=\frac{\gamma_{12}\lambda}{2}e^{-\lambda|\tau'-\tau|-{\rm i}\omega_{0}(\tau'-\tau)}$$ in the second equality of Eq.(30).

The double integrals in equations (28)-(30) can be further solved. Using
$$|\tau'-\tau|=\{\begin{array}{ccc}
                   \tau'-\tau & if & \tau'\geq\tau, \\
                   \tau-\tau' & if & \tau'<\tau,
                 \end{array}
$$ to divide the integral with respect to $\tau'$ into two parts, after tedious but straightforward calculations, we finally obtain
\begin{subnumcases}{}
  \sum_{k}|c_{k}(t)|^{2} =\alpha_{1}(t)|c_{3}(0)|^{2},\\
  \sum_{k}|d_{k}(t)|^{2} =\alpha_{2}(t)|c_{3}(0)|^{2},\\
  \sum_{k}c_{k}(t)d_{k}^{*}(t) =\Theta(t)|c_{3}(0)|^{2},
\end{subnumcases}
where
\begin{widetext}
\begin{small}
\begin{equation}
 \alpha_{1}(t)=\frac{\gamma_{1}\lambda}{2}\sum_{j,l=1}^{3}\mathfrak{D}_{j}\mathfrak{D}_{l}^{*}\left\{\Omega_{1}^{jl}e^{(\mathfrak{b}_{j}+\mathfrak{b}_{l}^{*})t}+\Omega_{2}^{jl}e^{(-\lambda+\mathfrak{b}_{j}+{\rm i}\delta_{1})t}+\Omega_{3}^{jl}e^{(-\lambda+\mathfrak{b}_{l}^{*}-{\rm i}\delta_{1})t}+\Omega_{4}^{jl}\right\},
\end{equation}
\begin{equation}
 \Theta(t)=\frac{\gamma_{12}\lambda}{2}\sum_{j,l=1}^{3}\mathfrak{D}_{j}\mathfrak{D}_{l}^{*}\left\{W_{1}^{jl}e^{(\mathfrak{b}_{j}+\mathfrak{b}_{l}^{*})t}+W_{2}^{jl}e^{(-\lambda+\mathfrak{b}_{j}+{\rm i}\delta_{2})t}+W_{3}^{jl}e^{(-\lambda+\mathfrak{b}_{l}^{*}-{\rm i}\delta_{1})t}+W_{4}^{jl}e^{{\rm i}(\omega_{1}-\omega_{2})t}\right\},
\end{equation}
with
\begin{eqnarray}
  \nonumber \Omega_{1}^{jl} &=& \frac{1}{(-\lambda+\mathfrak{b}_{l}^{*}-{\rm i}\delta_{1})(\lambda+\mathfrak{b}_{j}+{\rm i}\delta_{1})}
    - \frac{2\lambda}{(\lambda+\mathfrak{b}_{l}^{*}-{\rm i}\delta_{1})(-\lambda+\mathfrak{b}_{l}^{*}-{\rm i}\delta_{1})(\mathfrak{b}_{j}+\mathfrak{b}_{l}^{*})},\\
  \nonumber \Omega_{2}^{jl}& =& \frac{1}{(\lambda-\mathfrak{b}_{j}-{\rm i}\delta_{1})(\lambda+\mathfrak{b}_{l}^{*}-{\rm i}\delta_{1})},\\
  \nonumber W_{1}^{jl} &=& \frac{1}{(-\lambda+\mathfrak{b}_{l}^{*}-{\rm i}\delta_{2})(\lambda+\mathfrak{b}_{j}+{\rm i}\delta_{1})}
    - \frac{2\lambda}{(\lambda+\mathfrak{b}_{l}^{*}-{\rm i}\delta_{2})(-\lambda+\mathfrak{b}_{l}^{*}-{\rm i}\delta_{2})[\mathfrak{b}_{j}+\mathfrak{b}_{l}^{*}+{\rm i}(\omega_{2}-\omega_{1})]},\\
  \nonumber W_{2}^{jl} &=& \frac{1}{(\lambda+\mathfrak{b}_{l}^{*}-{\rm i}\delta_{2})(\lambda-\mathfrak{b}_{j}-{\rm i}\delta_{1})}.
\end{eqnarray}
\end{small}
\end{widetext}
The other four coefficients can be obtained through the following replacement
$$
\begin{array}{cccccc}
  \Omega_{1}^{jl} & \underrightarrow{~~\lambda\rightarrow-\lambda~~}& \Omega_{4}^{jl},&~~ \Omega_{2}^{jl} & \underrightarrow{~~\lambda\rightarrow-\lambda~~}& \Omega_{3}^{jl},
\end{array}
$$
$$
\begin{array}{cccccc}
  W_{1}^{jl} & \underrightarrow{~~\lambda\rightarrow-\lambda~~}& W_{4}^{jl},&~~ W_{2}^{jl} & \underrightarrow{~~\lambda\rightarrow-\lambda~~}& W_{3}^{jl}.
\end{array}
$$
In addition, the coefficient $\alpha_{2}(t)$ in equation (31b) can be obtained by making the replacement
\begin{equation}
\begin{array}{ccc}
  \alpha_{1}(t) & \underrightarrow{~~\gamma_{1}\rightarrow \gamma_{2},\omega_{1}\rightarrow \omega_{2}~~}& \alpha_{2}(t).
\end{array}
\end{equation}

Finally, the reduced density matrix of equation (21) can be written as
\begin{widetext}
\begin{equation}
    \varrho_{s}(t)=\left[
                        \begin{array}{ccc}
                          |c_{1}(0)|^{2}+\alpha_{1}(t)|c_{3}(0)|^{2} & c_{1}(0)c_{2}^{*}(0)e^{{\rm i}(\omega_{1}-\omega_{2})t}+\Theta(t)|c_{3}(0)|^{2} & T_{1}(t)c_{1}(0)c_{3}^{*}(0) \\
                          c_{2}(0)c_{1}^{*}(0)e^{{\rm i}(\omega_{2}-\omega_{1})t}+\Theta^{*}(t)|c_{3}(0)|^{2} & |c_{2}(0)|^{2}+\alpha_{2}(t)|c_{3}(0)|^{2} & T_{2}(t)c_{2}(0)c_{3}^{*}(0) \\
                          T_{1}^{*}(t)c_{3}(0)c_{1}^{*}(0) & T_{2}^{*}(t)c_{3}(0)c_{2}^{*}(0) & \alpha_{3}(t)|c_{3}(0)|^{2} \\
                        \end{array}
                      \right],
\end{equation}
\end{widetext}
where
\begin{eqnarray}
\nonumber T_{1}(t)&=&e^{{\rm i}\omega_{1}t}\left(\sum_{j=1}^{3}\mathfrak{D}_{j}^{*}e^{\mathfrak{b}_{j}^{*}t}\right),\\
\nonumber T_{2}(t)&=&e^{{\rm i}\omega_{2}t}\left(\sum_{j=1}^{3}\mathfrak{D}_{j}^{*}e^{\mathfrak{b}_{j}^{*}t}\right),\\
\nonumber \alpha_{3}(t)&=&\left|\sum_{j=1}^{3}\mathfrak{D}_{j}e^{\mathfrak{b}_{j}t}\right|^{2},
\end{eqnarray}
and $\alpha_{1}(t)$, $\alpha_{2}(t)$, $\Theta(t)$ are given by eqs.(32)-(33) and the replacement (34).

\section{Quantum interference}
Quantum interference is a kind of unique phenomenon of multilevel atomic systems different from two-level systems. The transitions from different channels may take place interference, leading to such as the well-known phenomenon of electromagnetically induced transparency (EIT). As the first example of applications of the analytical solutions presented in the above section, we discuss in this section the problem of quantum interference in the process of spontaneous emissions.

For the open V-type three-level atom, by observing the structure of $E(t)$, $F(t)$, $G(t)$, $H(t)$ in Eqs.(17a)-(17b), we find that the boundedness of $|C_{1}(t)|^{2}$ and $|C_{2}(t)|^{2}$ in the limit $t\rightarrow \infty$ implies that each of the roots $b_{i}$ ($i=1,2,3$) of the equation $p^{3}+h_{1}p^{2}+h_{2}p+h_{3}=0$ should have non-positive real part. When one or more of the roots have zero real parts (i.e.,pure imaginary roots), $|C_{1}(t)|^{2}$ and $|C_{2}(t)|^{2}$ may have nonzero asymptotic values. Otherwise, the asymptotic values are zero. The occurrence of the nonzero asymptotic values can be regarded as the result of quantum interference between the transitions $|1\rangle\rightarrow |0\rangle$ and $|2\rangle\rightarrow |0\rangle$, as explained in the following.

By setting the ansatz $p=i\chi$ into the equation $p^{3}+h_{1}p^{2}+h_{2}p+h_{3}=0$, we find that the necessary condition of pure imaginary root is $\omega_{1}=\omega_{2}
$. This is just one of the necessary conditions of optical interference. The another interference condition, i.e., parallel polarization, has already been guaranteed by the condition $\gamma_{12}=\gamma_{21}=\sqrt{\gamma_{1}\gamma_{2}}$. Under these conditions, can one observe the quantum interference?

In Fig.2, we plot the time evolution of the populations $|c_{1}(t)|^{2}$ and $|c_{2}(t)|^{2}$ under the above interference conditions, where we set $\omega_{1}=\omega_{2}=\omega_{0}=20\gamma$, $\lambda=2\gamma$. In Fig.2a and 2b, we choose $\gamma_{1}=\gamma_{2}\equiv\gamma$ and $c_{1}(0)=-c_{2}(0)$ (Fig.2a) or $c_{1}(0)=c_{2}(0)$ (Fig.2b). We find that the populations $|c_{1}(t)|^{2}$ and $|c_{2}(t)|^{2}$ keep almost unchanged when $c_{1}(0)$ and $c_{2}(0)$ have opposite signs, and reduce quickly to nearly zero when they have same signs. This is actually the result of destructive interference and constructive interference. The opposite signs between $c_{1}(0)$ and $c_{2}(0)$ means opposite phases between the transitions $|1\rangle\rightarrow |0\rangle$ and $|2\rangle\rightarrow |0\rangle$, leading to destructive interference of the transitions which prohibits the decaying of the populations $|c_{1}(t)|^{2}$ and $|c_{2}(t)|^{2}$. On the contrary, same signs between $c_{1}(0)$ and $c_{2}(0)$ leads to constructive interference which speeds up the decaying of the populations.

In Fig.2c and 2d, we let $\gamma_{1}\neq\gamma_{2}$ but satisfying $\gamma_{1}|c_{1}(0)|^{2}=\gamma_{2}|c_{2}(0)|^{2}$ (where $\gamma_{1}=\gamma$ and $\gamma_{2}=2\gamma, 3\gamma, 4\gamma$ for the blue, red, black lines respectively). It is shown similar results: The populations $|c_{1}(t)|^{2}$ and $|c_{2}(t)|^{2}$ keep almost unchanged when $c_{1}(0)$ and $c_{2}(0)$ have opposite signs, and reduce quickly when they have same signs. This can be explained as follows: Though the decaying rates of the two excited states $\gamma_{1}$ and $\gamma_{2}$ are different, the condition $\gamma_{1}|c_{1}(0)|^{2}=\gamma_{2}|c_{2}(0)|^{2}$ guarantees that the decaying strengthes from two transitions are same. Thus the destructive interference in Fig.2c is complete and the populations can maintain unchanged. On the contrary, for the case of $\gamma_{1}|c_{1}(0)|^{2}\neq\gamma_{2}|c_{2}(0)|^{2}$ (Fig.2e and 2f, where $\gamma_{1}=\gamma$ and $\gamma_{2}=3\gamma, 4\gamma, 2\gamma$ for the blue, red, black lines respectively), the difference of the decaying strengthes from the two transitions gives rise to incompletely destructive interference, leading the the change of populations. As the light emitted from the higher-strength decaying channel has excess part which can excite in turn the lower-strength channel, leading to the increase of asymptotic population that corresponds to lower-strength decaying channel (Fig.2e). In addition, for the constructive interference, as the decaying rates $\gamma_{1}$ and $\gamma_{2}$ in Fig.2b are small, the energy lost in the environment spreads out timely, leading to the monotonic decrease of the populations. However in Fig.2d and Fig.2f, with the increasing of $\gamma_{2}$, the energy lost in the environment can not spread out timely, leading to the re-excitation of the populations. This is actually the commonly so-called memory effect.

\begin{figure}
  \begin{minipage}[t]{0.5\linewidth}
    \includegraphics[width=1.7in,height=4.3cm]{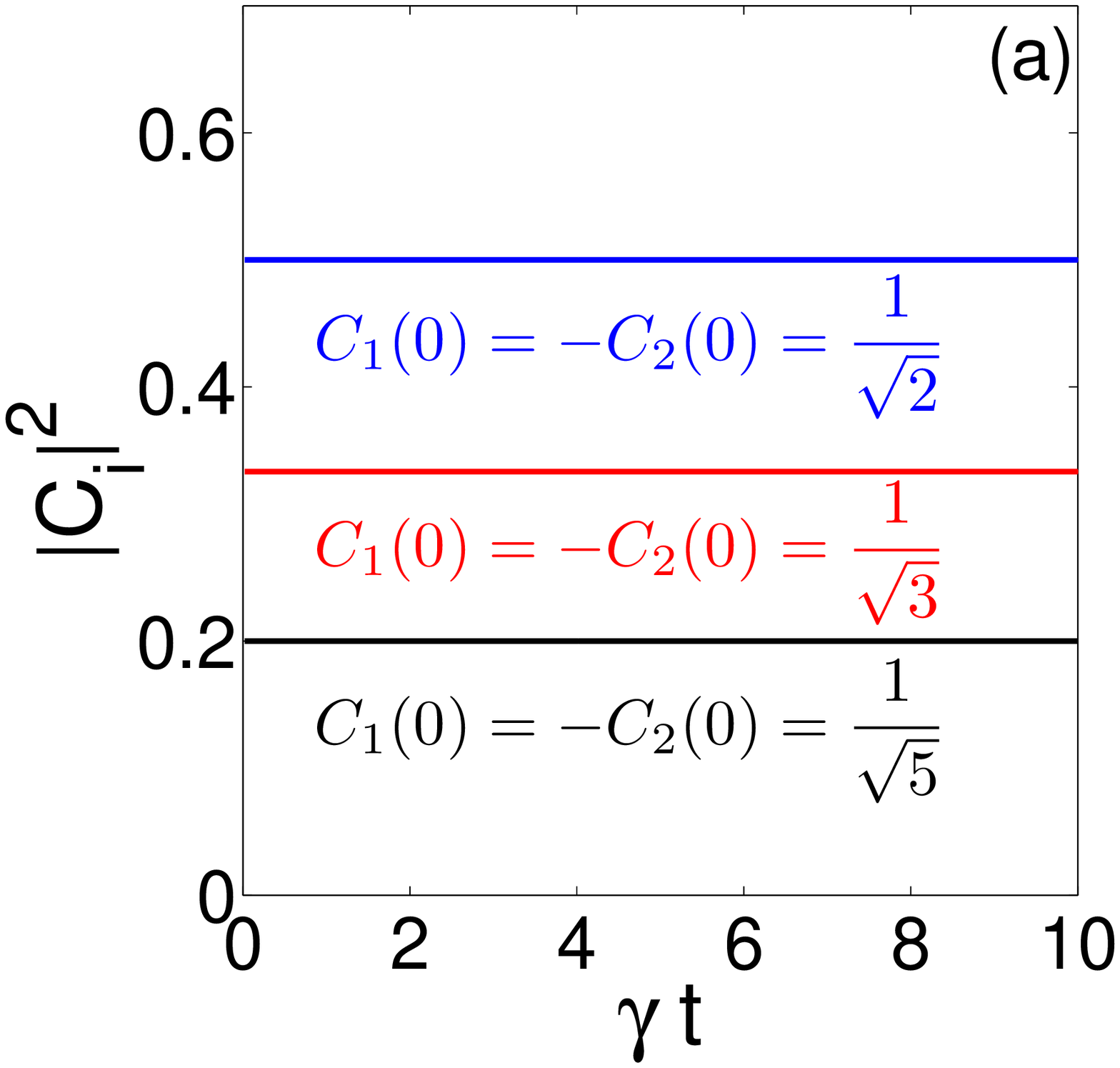}
  \end{minipage}%
    \begin{minipage}[t]{0.5\linewidth}
    \includegraphics[width=1.7in,height=4.3cm]{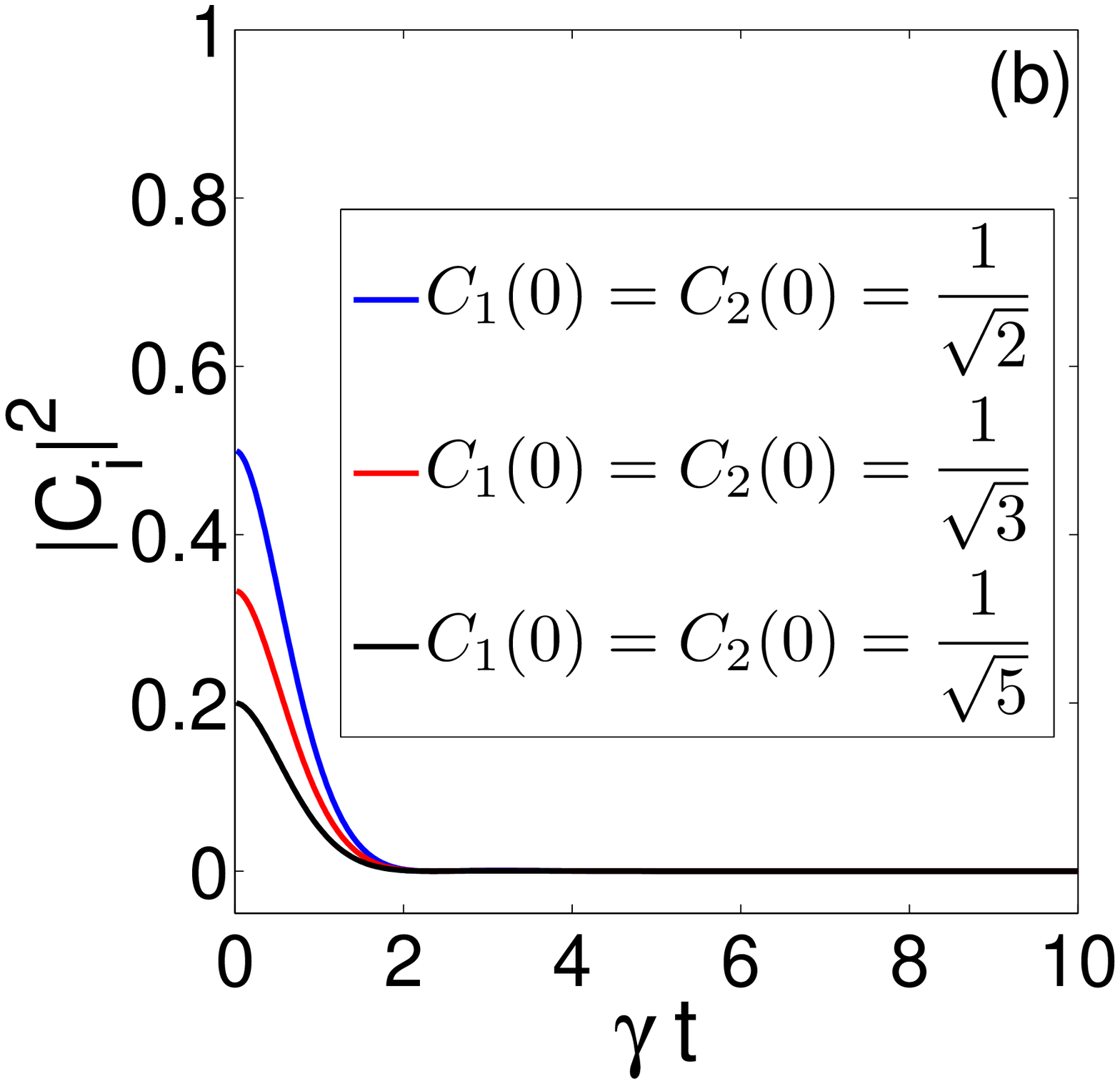}
  \end{minipage}
  \begin{minipage}[t]{0.5\linewidth}
    \includegraphics[width=1.7in,height=4.3cm]{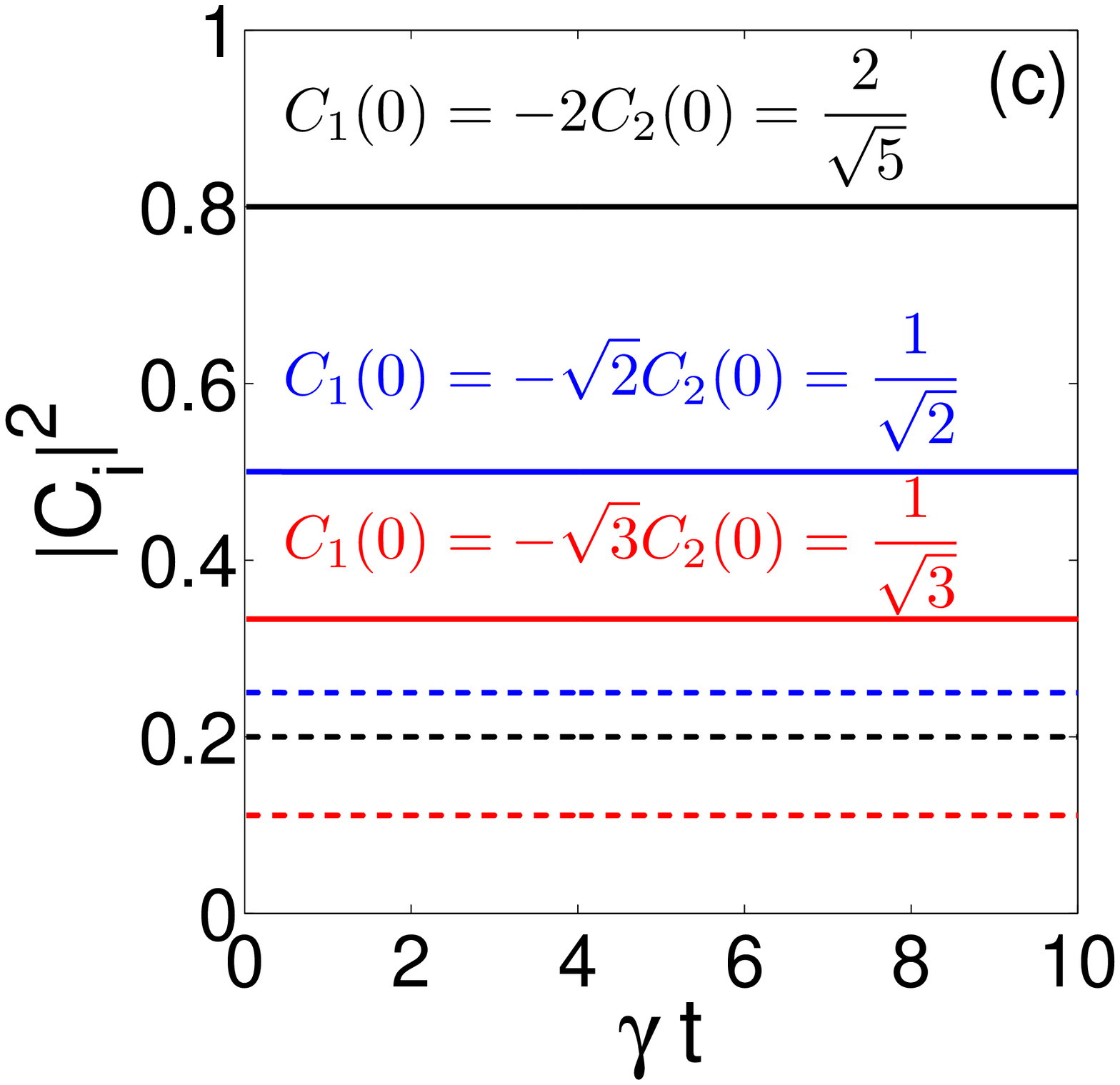}
  \end{minipage}%
    \begin{minipage}[t]{0.5\linewidth}
    \includegraphics[width=1.7in,height=4.3cm]{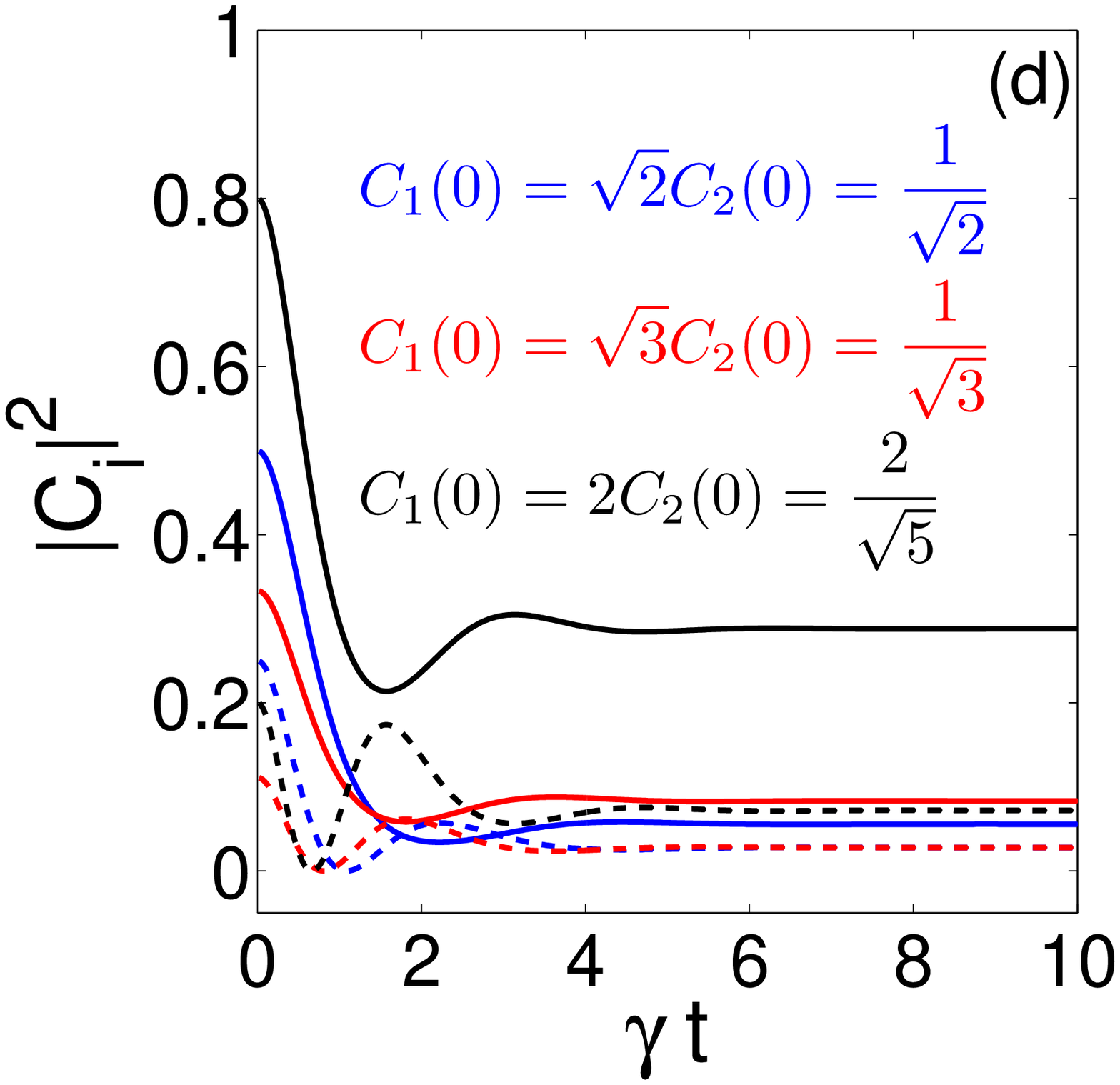}
  \end{minipage}
  \begin{minipage}[t]{0.5\linewidth}
    \includegraphics[width=1.7in,height=4.3cm]{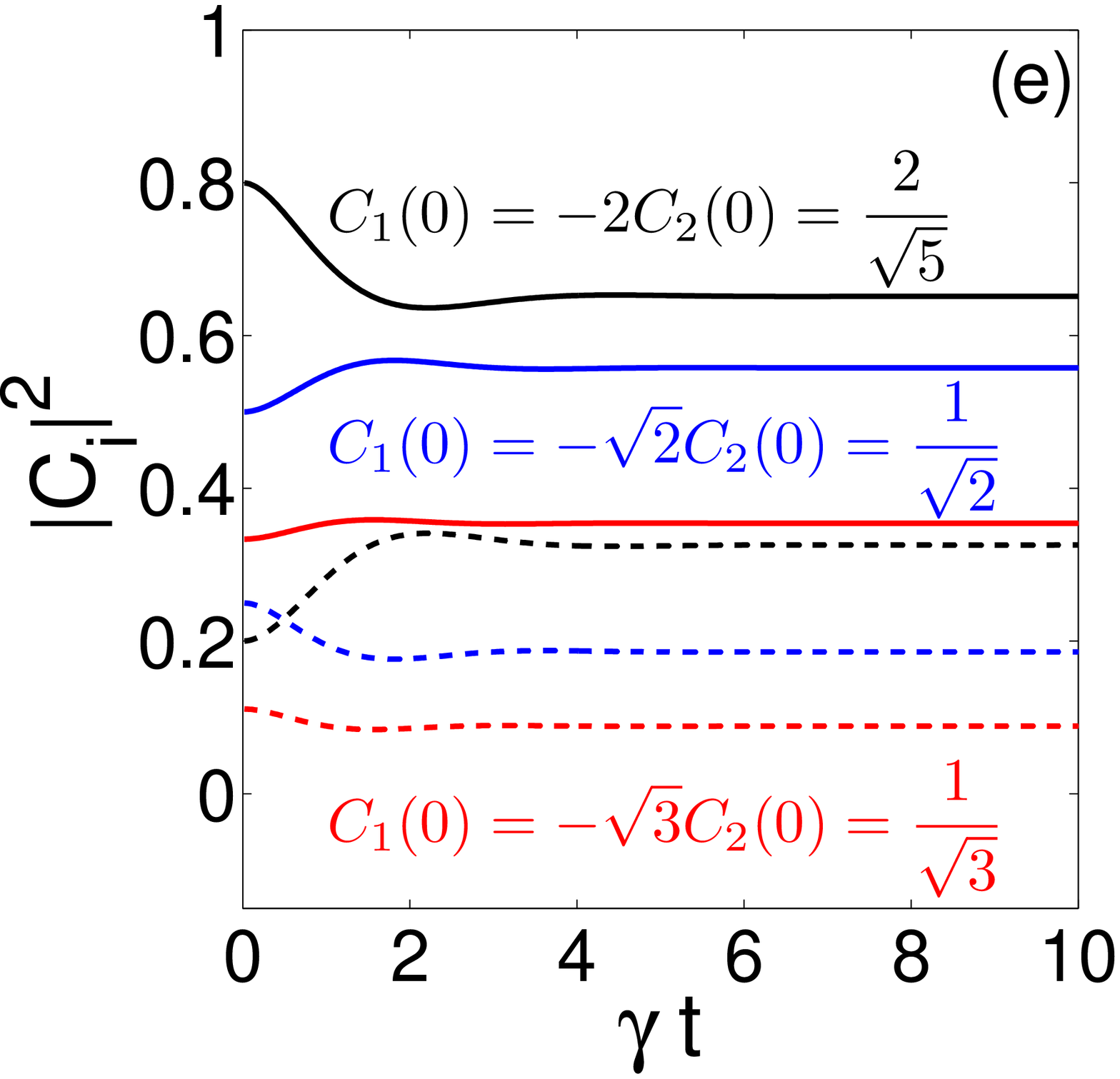}
  \end{minipage}%
    \begin{minipage}[t]{0.5\linewidth}
    \includegraphics[width=1.7in,height=4.3cm]{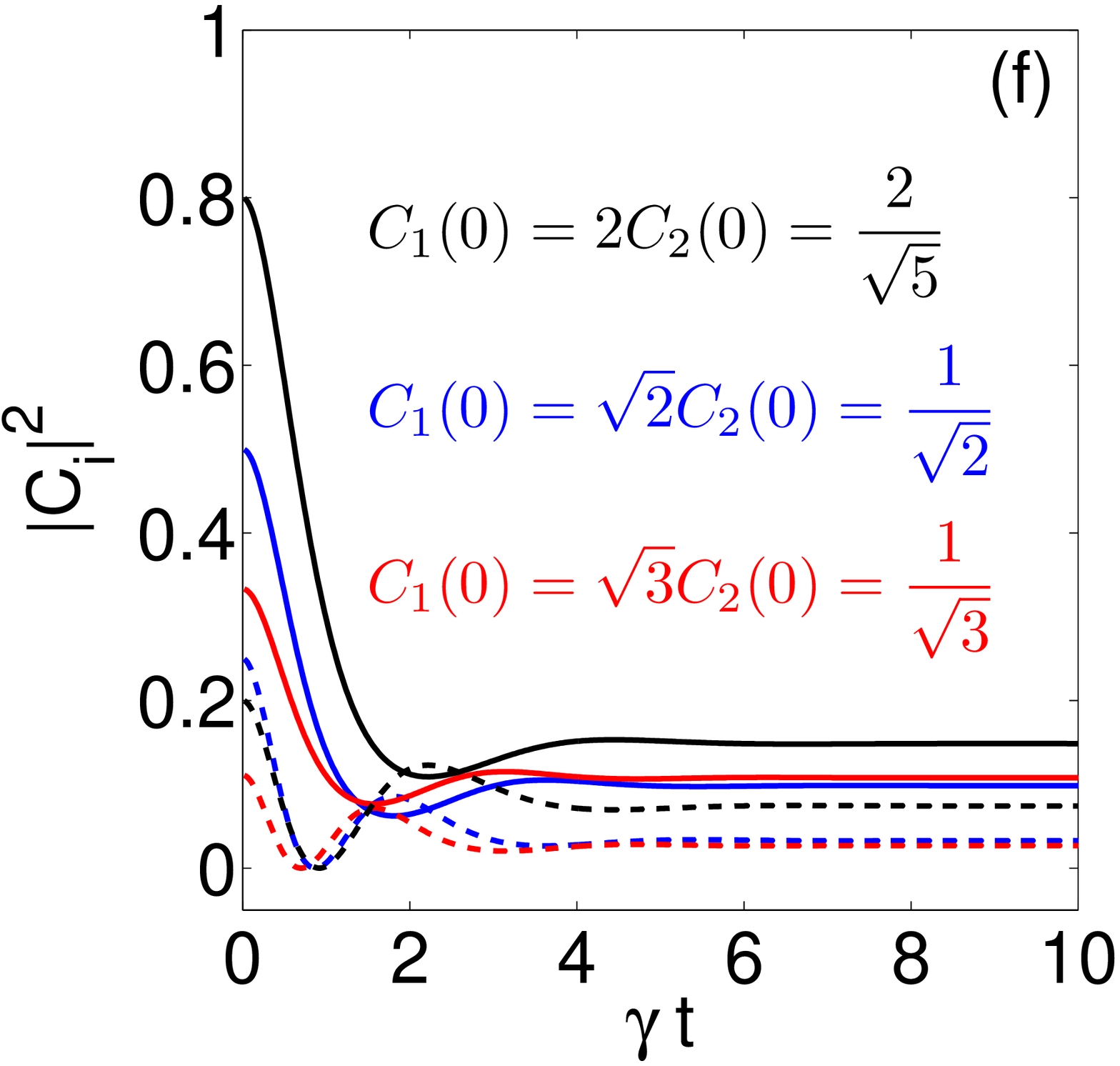}
  \end{minipage}
  \caption{\small Evolution of the populations versus dimensionless time $\gamma t$ for the open V-type atom with initial state given by Eq.(2), where $\omega_{1}=\omega_{2}=\omega_{0}=20\gamma$, $\lambda=2\gamma$. The solid lines denote $|c_{1}(t)|^{2}$ and dash lines denote $|c_{2}(t)|^{2}$. The colors of the curves correspond to respectively the initial states of the corresponding colors. (a) and (b) $\gamma_{1}=\gamma_{2}=\gamma$; (c) and (d) $\gamma_{1}|c_{1}(0)|^{2}=\gamma_{2}|c_{2}(0)|^{2}$; (e) and (f) $\gamma_{1}|c_{1}(0)|^{2}\neq\gamma_{2}|c_{2}(0)|^{2}$.}
\end{figure}

\begin{figure}
\includegraphics[width=3.5in,height=6cm]{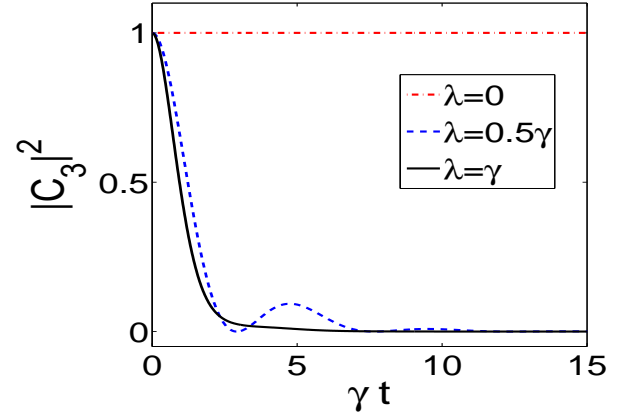}
\caption{ \small Time evolution of the population of the upper-level state for the $\Lambda$-type atom. Where we set $\gamma_{1}=\gamma_{2}=\gamma$ and the other parameters are $\lambda=0$, $\omega_{0}=91\gamma$, $\omega_{1}=\omega_{2}=90\gamma$ for the red dot-dash line; $\lambda=0.5\gamma$, $\omega_{1}=\omega_{2}=\omega_{0}=90\gamma$ for the blue dash line; $\lambda=\gamma$, $\omega_{1}=90\gamma$, $\omega_{2}=92\gamma$, $\omega_{0}=91\gamma$ for the black solid line.}
\end{figure}

We can also discuss in a similar way the quantum interference of $\Lambda$-type atom in the process of spontaneous emissions. From Eq.(26), we see that the necessary condition for nonzero asymptotic population of the upper-level state of the $\Lambda$-type atom is that the real parts of $\mathfrak{b}_{i}$ must be zero. By setting ansatz $p=i\chi$ into the equation $R(p)=p^{3}+(M_{1}+M_{2})p^{2}+(M_{1}M_{2}+\frac{\gamma_{1}\lambda}{2}+\frac{\gamma_{2}\lambda}{2})p+\frac{\gamma_{1}\lambda}{2}M_{2}+\frac{\gamma_{2}\lambda}{2}M_{1}=0$, one obtains
\begin{subnumcases}{}
  2\chi^{2}+(\delta_{1}+\delta_{2})\chi-\frac{\lambda}{2}(\gamma_{1}+\gamma_{2}) = 0,\\
  \nonumber \chi^{3}+(\delta_{1}+\delta_{2})\chi^{2}-[\lambda^{2}-\delta_{1}\delta_{2}+\frac{\lambda}{2}(\gamma_{1}+\gamma_{2})]\chi\\
  ~~~~~~~~~~~~~~~~~~~~~-\frac{\lambda}{2}(\gamma_{1}\delta_{2}+\gamma_{2}\delta_{1}) = 0.
\end{subnumcases}
Though there are several adjustable structure parameters $\gamma_{j}$, $\delta_{j}$ and $\lambda$, this set of equations have not real root for $\chi$ when $\lambda\neq 0$ (see the proof in Appendix B). Thus the asymptotic population of the upper-level state must be zero when $t\rightarrow\infty$. Note that when $\lambda=0$ the set of equations have real roots $\chi=0$ and $\chi=-\delta_{1}$, with the former valid for any structure parameters and the latter valid for $\delta_{1}=\delta_{2}$. In fact, when $\lambda=0$, the correlation functions $f_{j}(t-\tau)$ ($j=1,2$) in Eq.(25) are zero. thus $|c_{3}(t)|^{2}$ remains its initial value unchanged. We plot the time evolution of the population $|c_{3}(t)|^{2}$
as in Fig.3 for several different set of structure parameters. The oscillation of the blue dash line originates from the non-Markovian effect. This
analysis suggests that when a three-level $\Lambda$-type atom takes place spontaneous emission in a real Lorentzian environment ($\lambda\neq 0$), there is no quantum interference between the two transition channels.

\section{Evolution of quantum Fisher information}
As the second example of applications, we study in this section the evolution of QFI of the parameters encoded in the open three-level atom. Owing to the absence of the quantum interference for the spontaneous emission of the $\Lambda$-type atom, we will take the V-type atom as an exemplum for investigation so as to highlight the roles of quantum interference. Now let us shortly review the notion of QFI. In quantum metrology, the problem of determining the optimal measurement scheme for a particular estimation scenario is non-trivial. Fortunately QFI provides us a useful tool for estimating the precision of a parameter measurement. The famous quantum Cram\'{e}r-Rao theorem, $\Delta^{2}\phi\geq 1/(\nu F_{\phi})$, presents the lower bound of the mean-square error of the unbiased estimator for the parameter $\phi$. Here $\nu$ denotes the number of repeated experiments and the QFI is defined through the symmetric logarithmic derivative as $F_{\phi}={\rm Tr}(\rho_{\phi}L_{\phi}^{2})$ with $\partial\rho_{\phi}/\partial\phi=(L_{\phi}\rho_{\phi}+\rho_{\phi}L_{\phi})/2$. By diagonalizing the density matrix as $\rho_{\phi}=\sum_{n}\lambda_{n}|\psi_{n}\rangle\langle\psi_{n}|$, one can write the QFI as\cite{Weizhong}
\begin{small}
\begin{equation}
    F_{\phi}=\sum_{n}\frac{(\partial_{\phi}\lambda_{n})^{2}}{\lambda_{n}}+2\sum_{n\neq m}\frac{(\lambda_{n}-\lambda_{m})^{2}}{\lambda_{n}+\lambda_{m}}|\langle\psi_{n}|\partial_{\phi}\psi_{m}\rangle|^{2},
\end{equation}
\end{small}
where the first and the second summations involve all sums but $\lambda_{n}\neq0$ and $\lambda_{n}+\lambda_{m}\neq 0$.

For the sake of convenience in the numerical simulations, we rewrite the superposition coefficients in Eq.(2) as
$c_{0}(0)=\cos\theta$, $c_{1}(0)=\sin\theta\sin\phi e^{{\rm i}\eta_{1}}$, $c_{2}(0)=\sin\theta\cos\phi e^{{\rm i}\eta_{2}}$. The simulations can be done through the Eqs. (4), (5a), (17a)-(17b) and (37). In Fig.4 and Fig.5, we plot the time evolution of QFI for the parameters $\theta$, $\phi$, $\eta_{1}$ and $\eta_{2}$ encoded in the open V-type atom. In the numerical simulations, we set $\gamma_{1}=\gamma_{2}=\gamma$, $\omega_{1}=90\gamma$, $\omega_{2}=92\gamma$ and $\omega_{0}=91\gamma$. It shows that the QFI decreases in general with time, implying the losing of the information from the open quantum system to the environment. For the smaller spectral width $\lambda$ (Fig.4), QFI decays slower and appears oscillations for relatively longer time. This is the result of the well-known memory effect presented by the non-Markovian dynamics. When the spectral width becomes larger (Fig.5), the memory effect becomes weaker so that QFI decays faster and the oscillations disappear gradually. When $\lambda=3\gamma$, the dynamics becomes basically Markovian.

\begin{figure}
  \begin{minipage}[t]{0.5\linewidth}
    \includegraphics[width=1.7in,height=5cm]{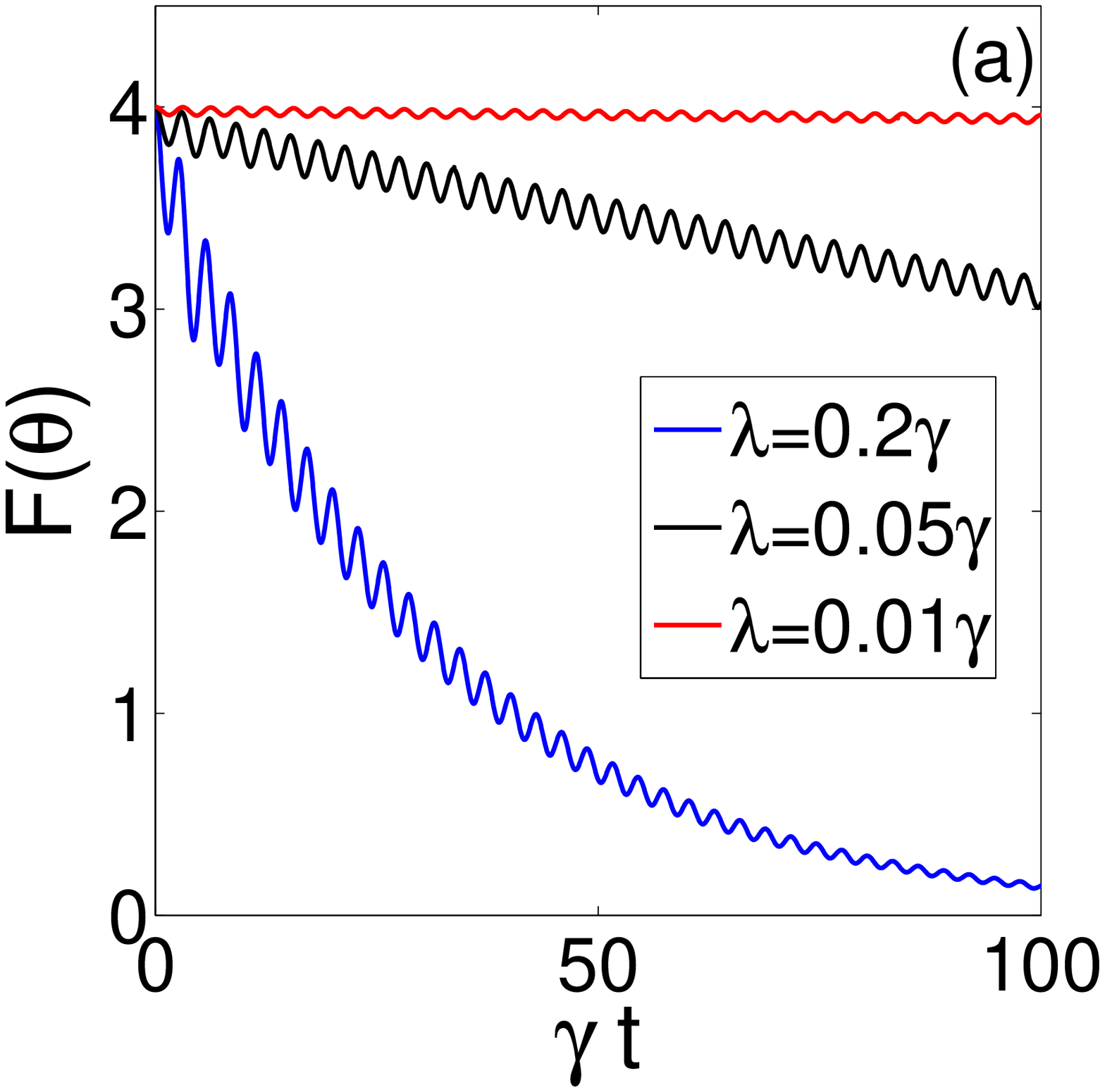}
  \end{minipage}%
  \hspace{-0.10in}
  \begin{minipage}[t]{0.5\linewidth}
    \includegraphics[width=1.7in,height=5cm]{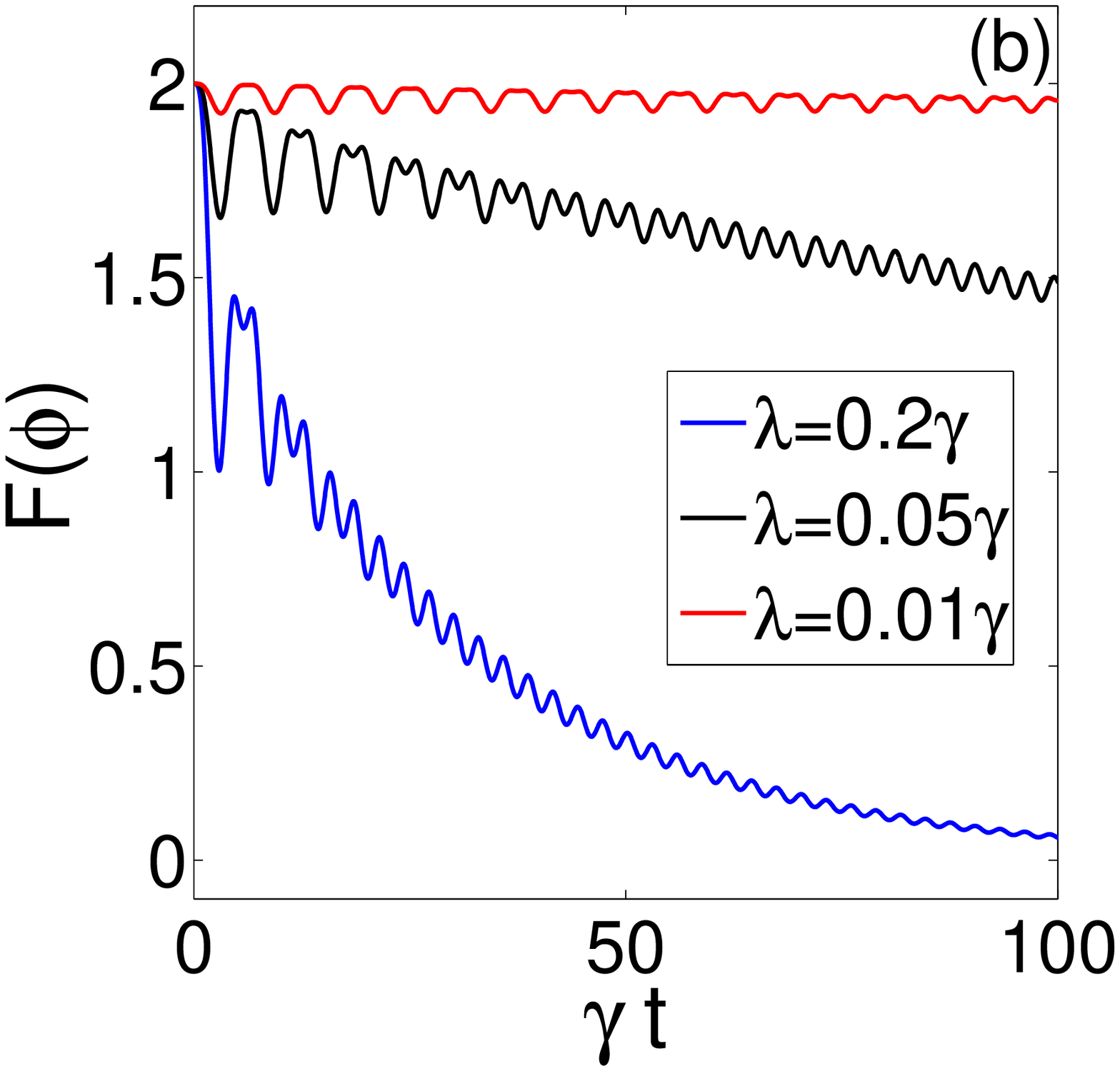}
  \end{minipage}
    \begin{minipage}[t]{0.5\linewidth}
    \includegraphics[width=1.7in,height=5cm]{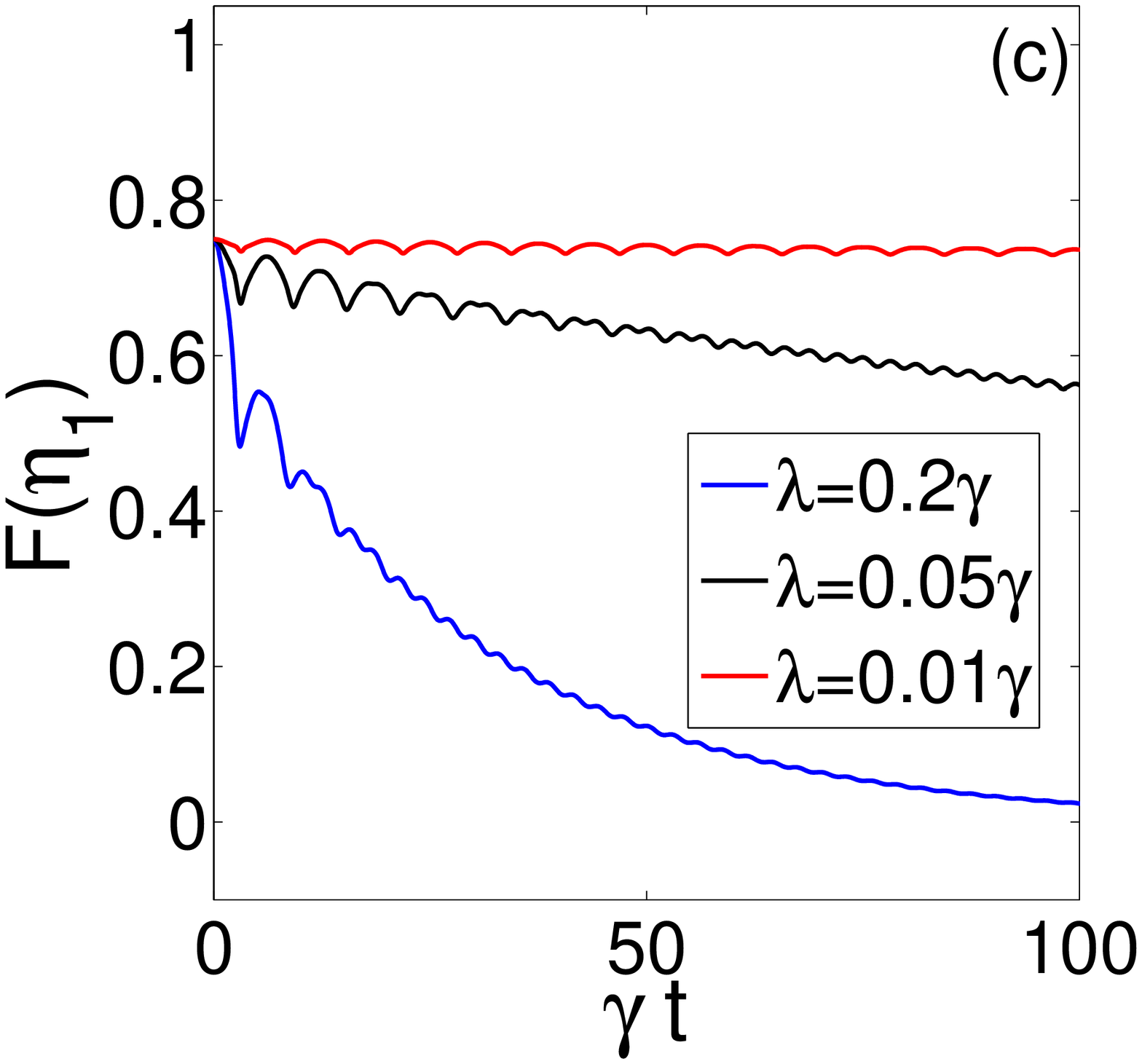}
  \end{minipage}%
    \hspace{-0.10in}
  \begin{minipage}[t]{0.5\linewidth}
    \includegraphics[width=1.7in,height=5cm]{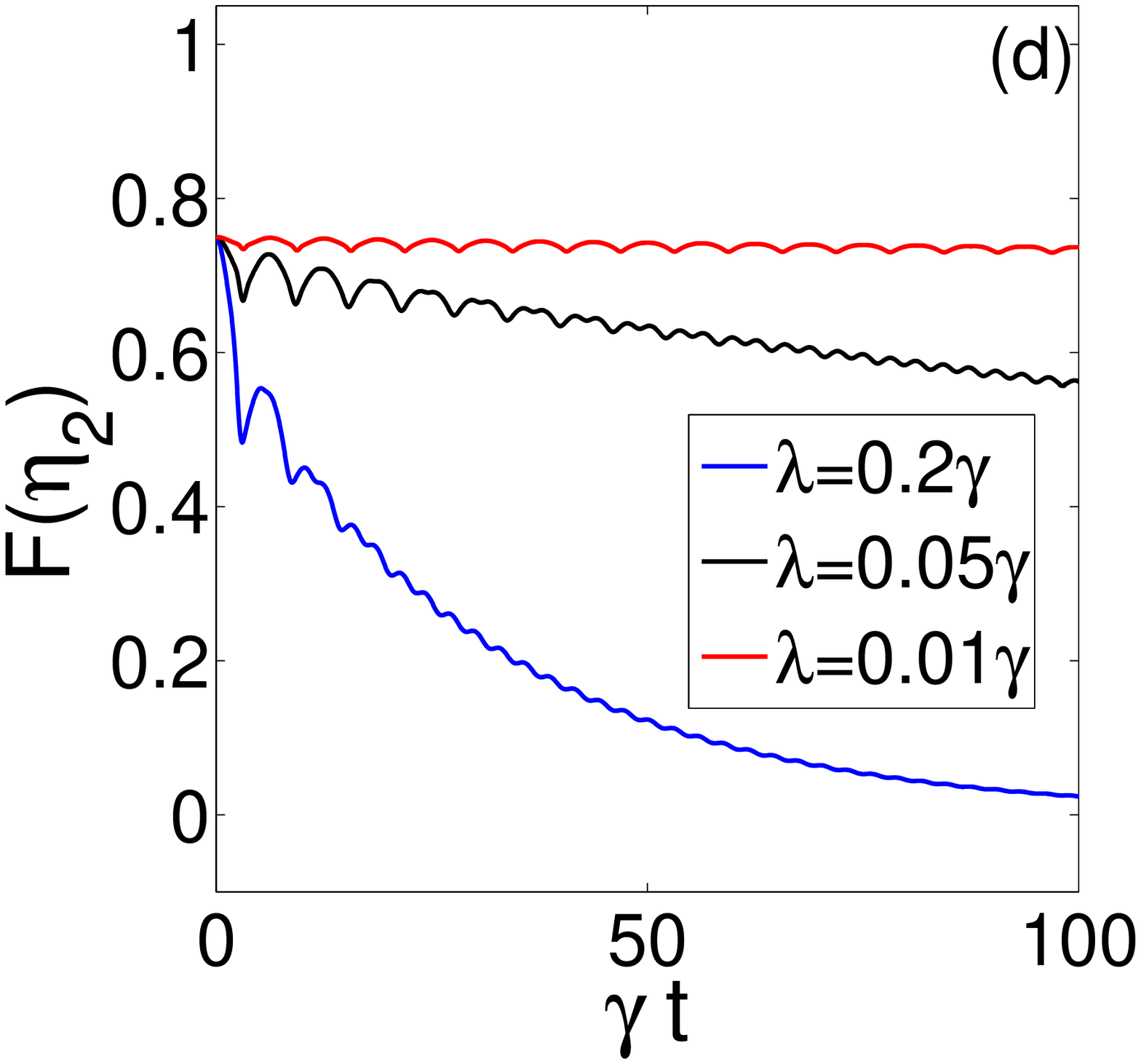}
  \end{minipage}
  \caption{\small Evolution of QFI versus dimensionless time $\gamma t$ for the parameters $\theta$, $\phi$, $\eta_{1}$ and $\eta_{2}$ for the open V-type atom. The parameters are chosen as $\gamma_{1}=\gamma_{2}=\gamma$, $\omega_{1}=90\gamma$, $\omega_{2}=92\gamma$ and $\omega_{0}=91\gamma$. The spectral widthes are displayed in the figure. The QFIs are evaluated at $\theta=\phi=\pi/4$, $\eta_{1}=\eta_{2}=\pi/2$.}
\end{figure}

\begin{figure}
  \vspace{-0.2cm}
  \begin{minipage}[t]{0.5\linewidth}
    \includegraphics[width=1.7in,height=5cm]{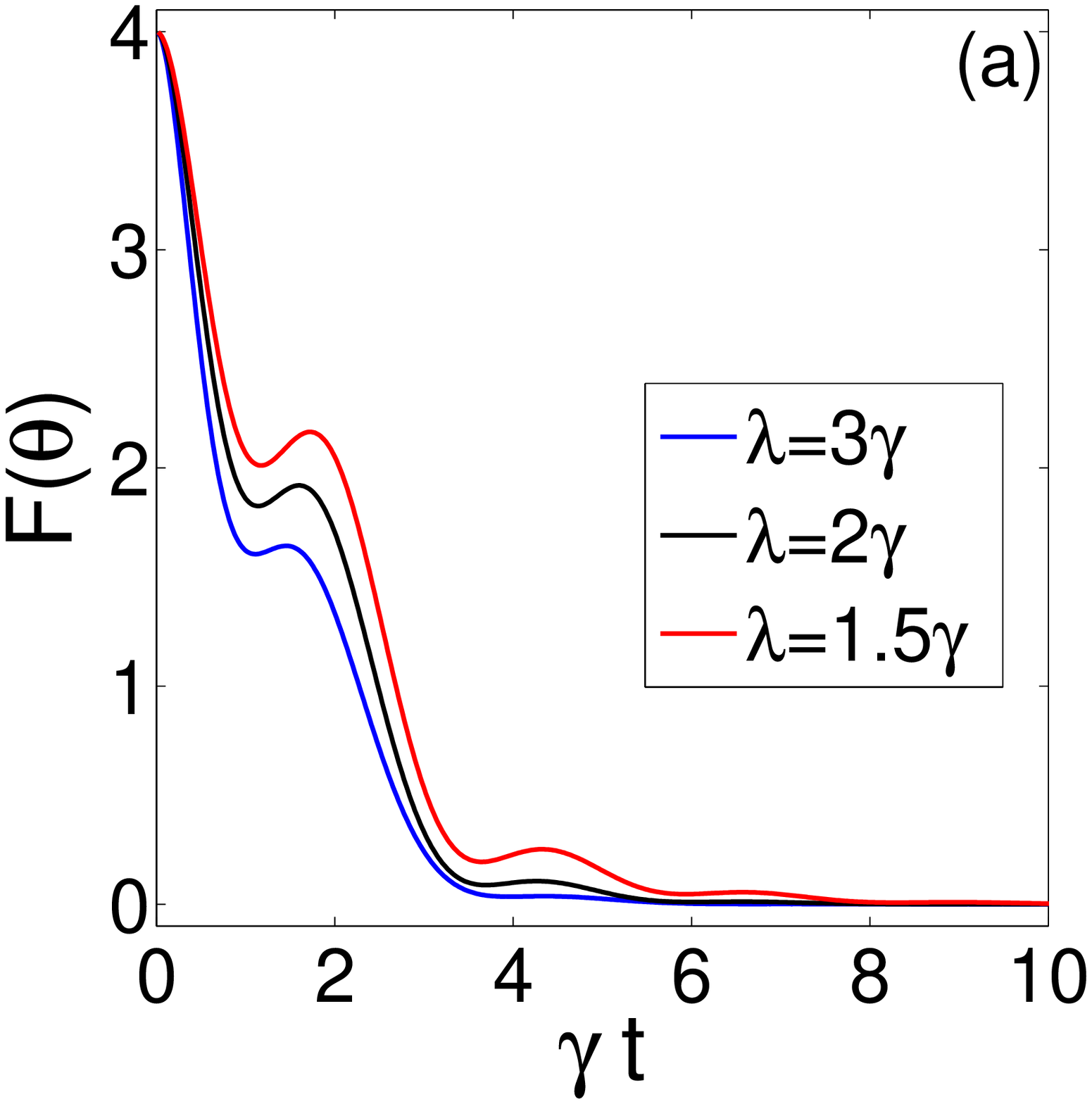}
  \end{minipage}%
  \hspace{-0.15in}
  \begin{minipage}[t]{0.5\linewidth}
    \includegraphics[width=1.7in,height=5cm]{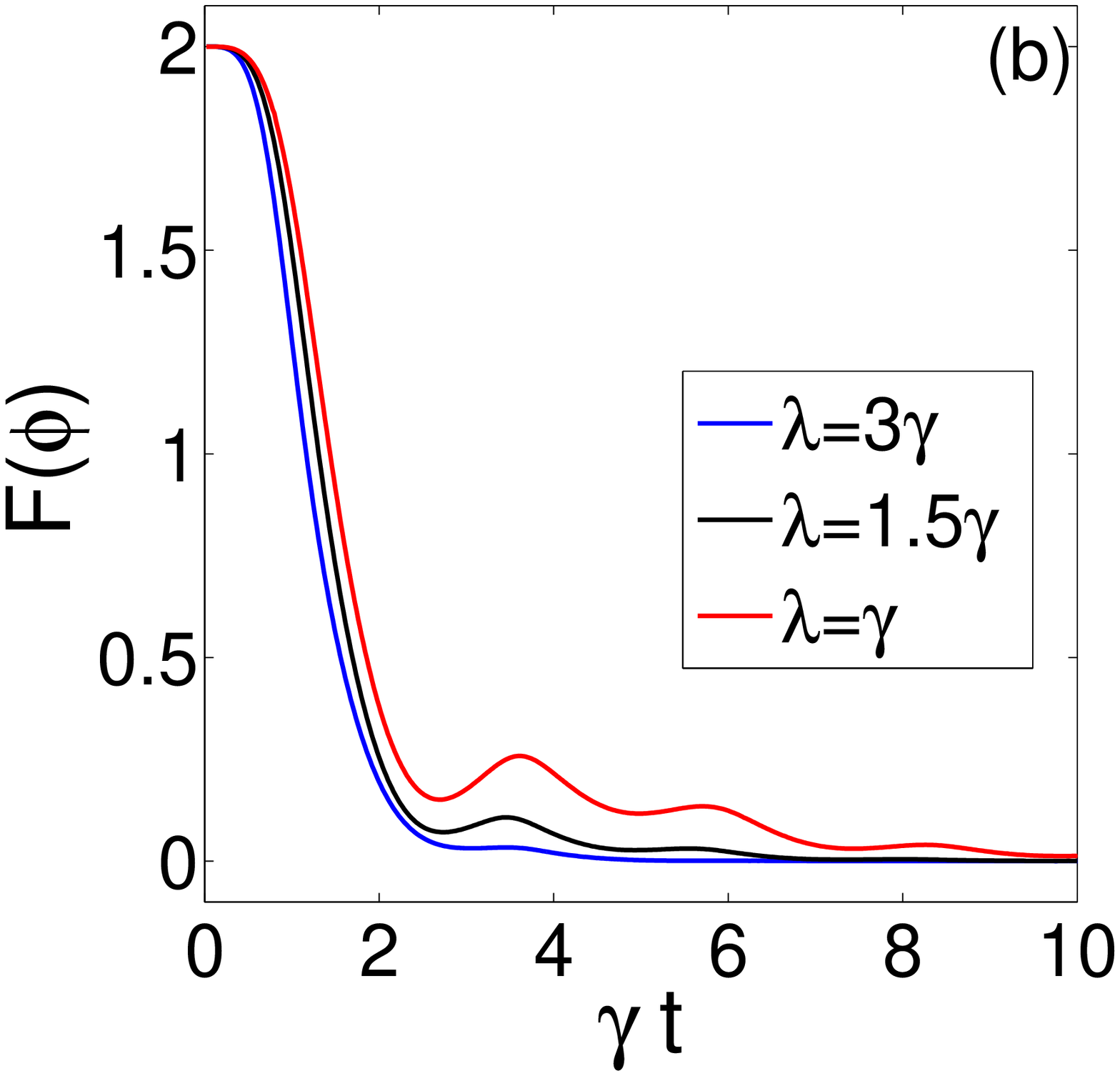}
  \end{minipage}
  \begin{minipage}[t]{0.5\linewidth}
    \includegraphics[width=1.7in,height=5cm]{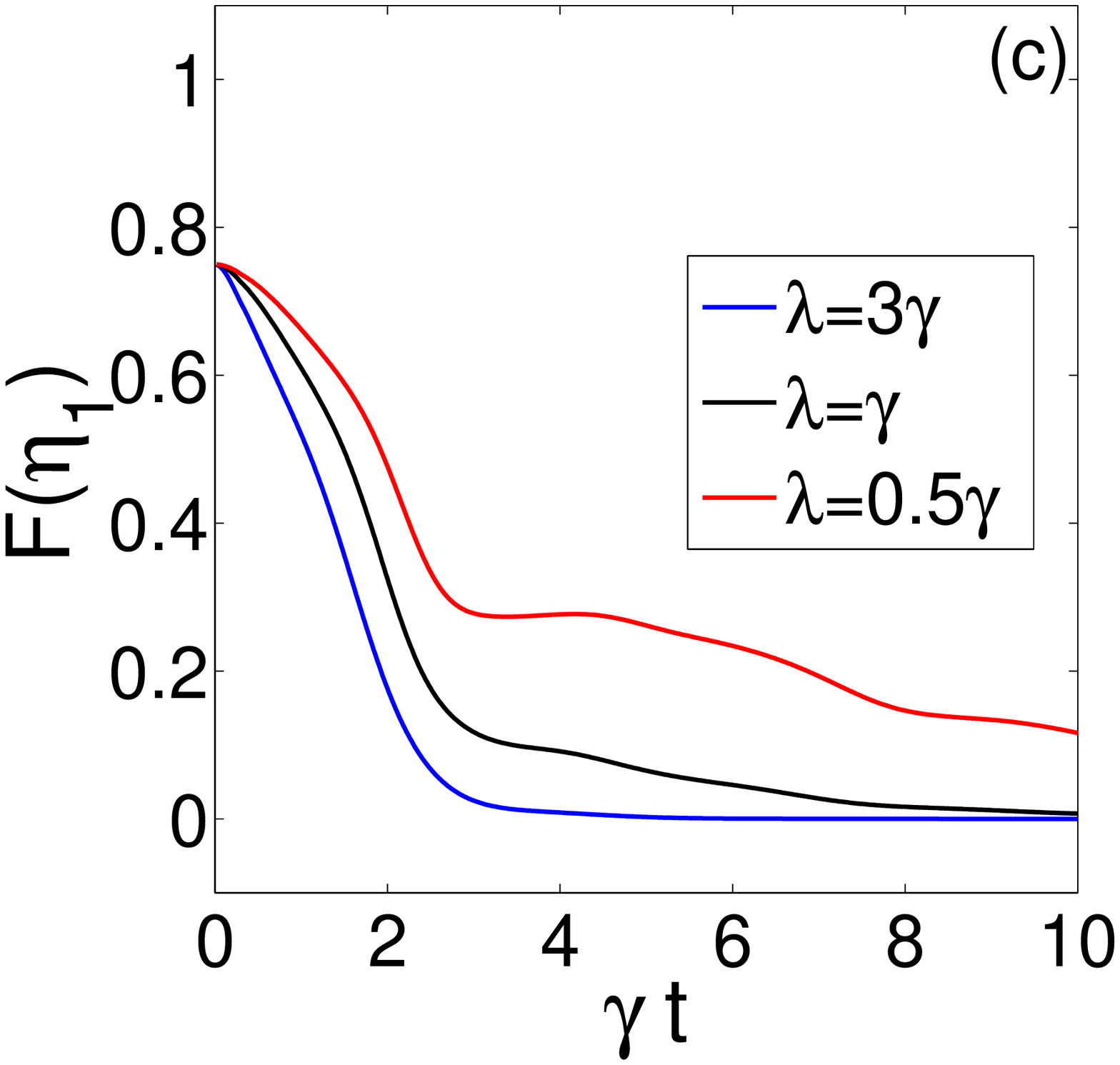}
  \end{minipage}%
  \hspace{-0.15in}
  \begin{minipage}[t]{0.5\linewidth}
    \includegraphics[width=1.7in,height=5cm]{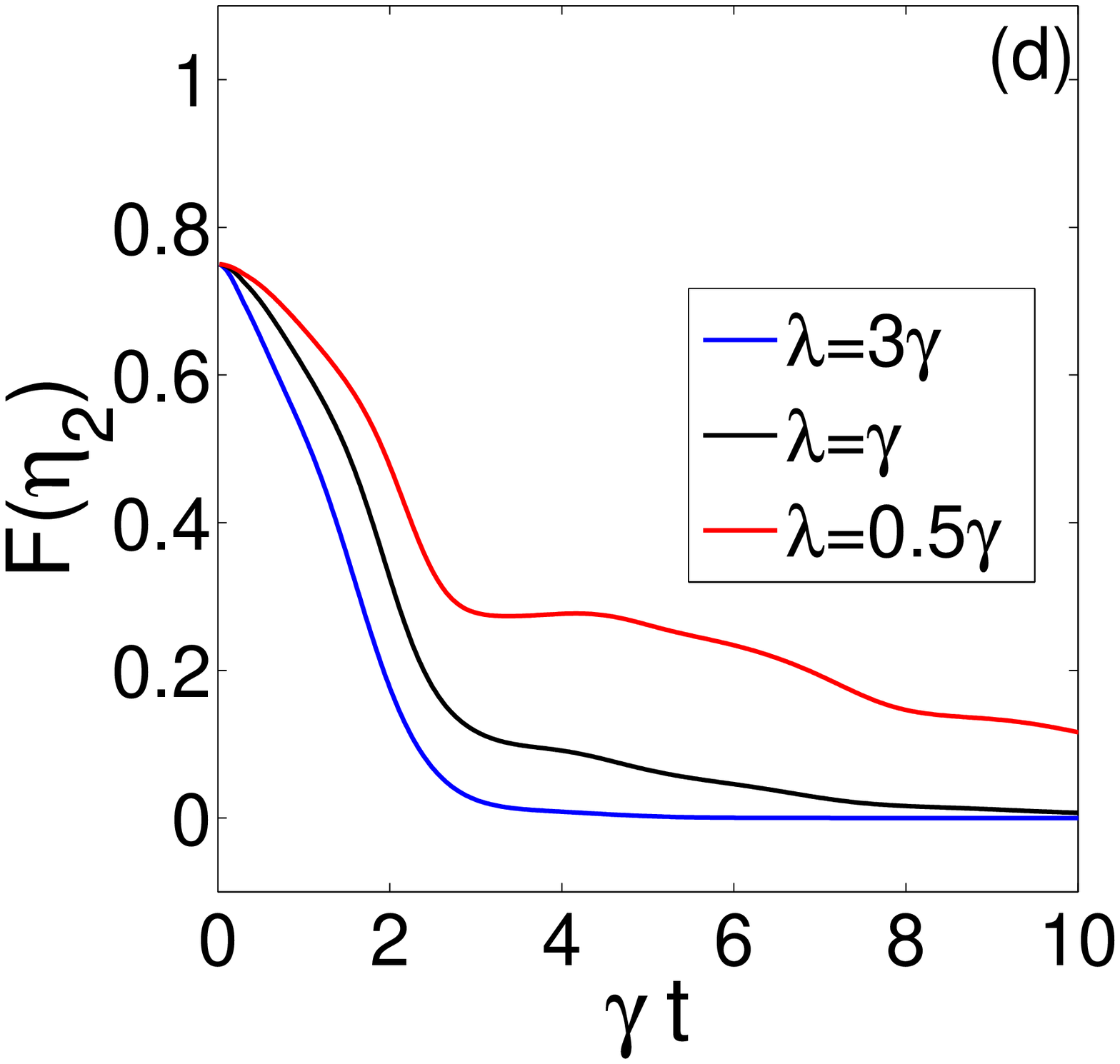}
  \end{minipage}
  \caption{\small Time evolution of QFI of the parameters $\theta$, $\phi$, $\eta_{1}$ and $\eta_{2}$ for the open V-type atom. Except for the different parameter $\lambda$ displayed in the figure, all other parameters are the same as in Fig.4.}
\end{figure}

An idea naturally arises: Whether can we use the quantum interference to protect quantum Fisher information? To answer this problem, we plot the time evolution of QFI under the interference conditions as in Fig.6, where $\gamma_{1}=\gamma_{2}=\gamma$, $\lambda=2\gamma$, $\omega_{1}=\omega_{2}=\omega_{0}=91\gamma$. The QFIs are evaluated at $\theta=\phi=\pi/4$, $\eta_{1}=\pi/2$, and $\eta_{2}=\pi/2$ (blue lines) or $\eta_{2}=3\pi/2$ (red lines). The blue lines correspond to cases of destructive interference and the red lines to the cases of constructive interference. It shows that QFI in some cases can really obtain good protection, but in other cases may decay more faster(compared Fig.6a and 6b with Fig.5a and 5b). Furthermore, destructive interference not always leads to the more effective protection to QFI, while the constructive interference is also not always detrimental. The reason is very simple. The destructive interference protects only the absolute values of $c_{1}(t)$ and $c_{2}(t)$, not themselves. The atomic state is thus changing in time. This approves actually the role of the relative phases in a quantum superposition state.

In Fig.7, we present the time evolution of QFI of the parameters for different values of the central frequency $\omega_{0}$ of the Lorentzian spectrum, where we set $\omega_{1}=90\gamma$, $\omega_{2}=92\gamma$, $\gamma_{1}=\gamma_{2}=\gamma$, and $\lambda=2\gamma$. It is shown that when $\omega_{0}$ locates at the middle between $\omega_{1}$ and $\omega_{2}$, i.e., $\omega_{0}=91\gamma$, the decaying of QFI is the fastest. Deviating from this middle frequency to the two sides, the decaying becomes slower and slower. Interestingly, the decaying of $F(\theta)$ and $F(\phi)$ seems to be symmetrical with respect to middle frequency $\omega_{0}=91\gamma$ as shown in Fig.7a and Fig.7b. This is inferred to be related to the symmetry of Lorentzian distribution, but the details remain to be confirmed.

\begin{figure}
  \begin{minipage}[t]{0.5\linewidth}
    \includegraphics[width=1.7in,height=4.3cm]{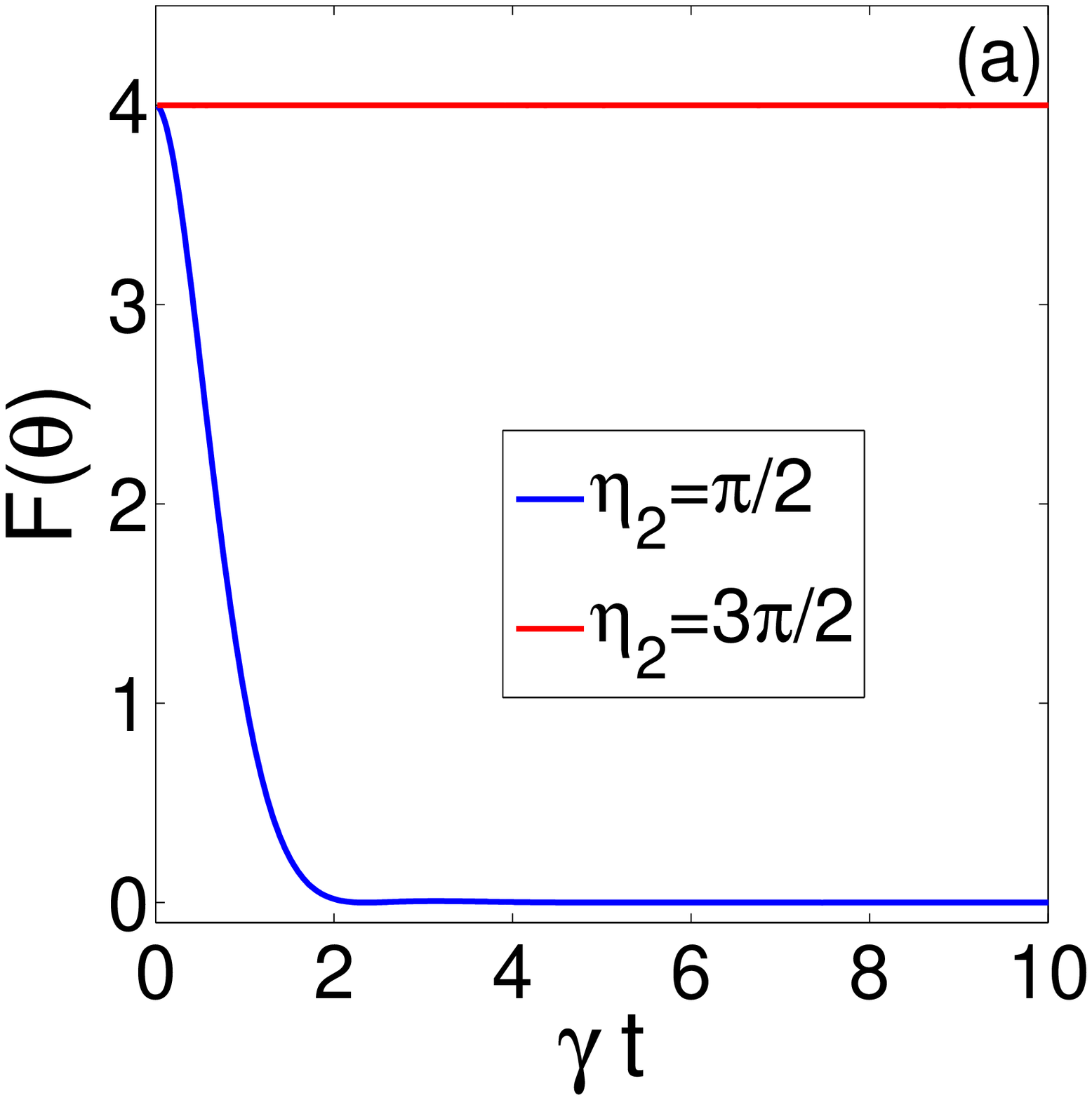}
  \end{minipage}%
  \begin{minipage}[t]{0.5\linewidth}
    \includegraphics[width=1.7in,height=4.3cm]{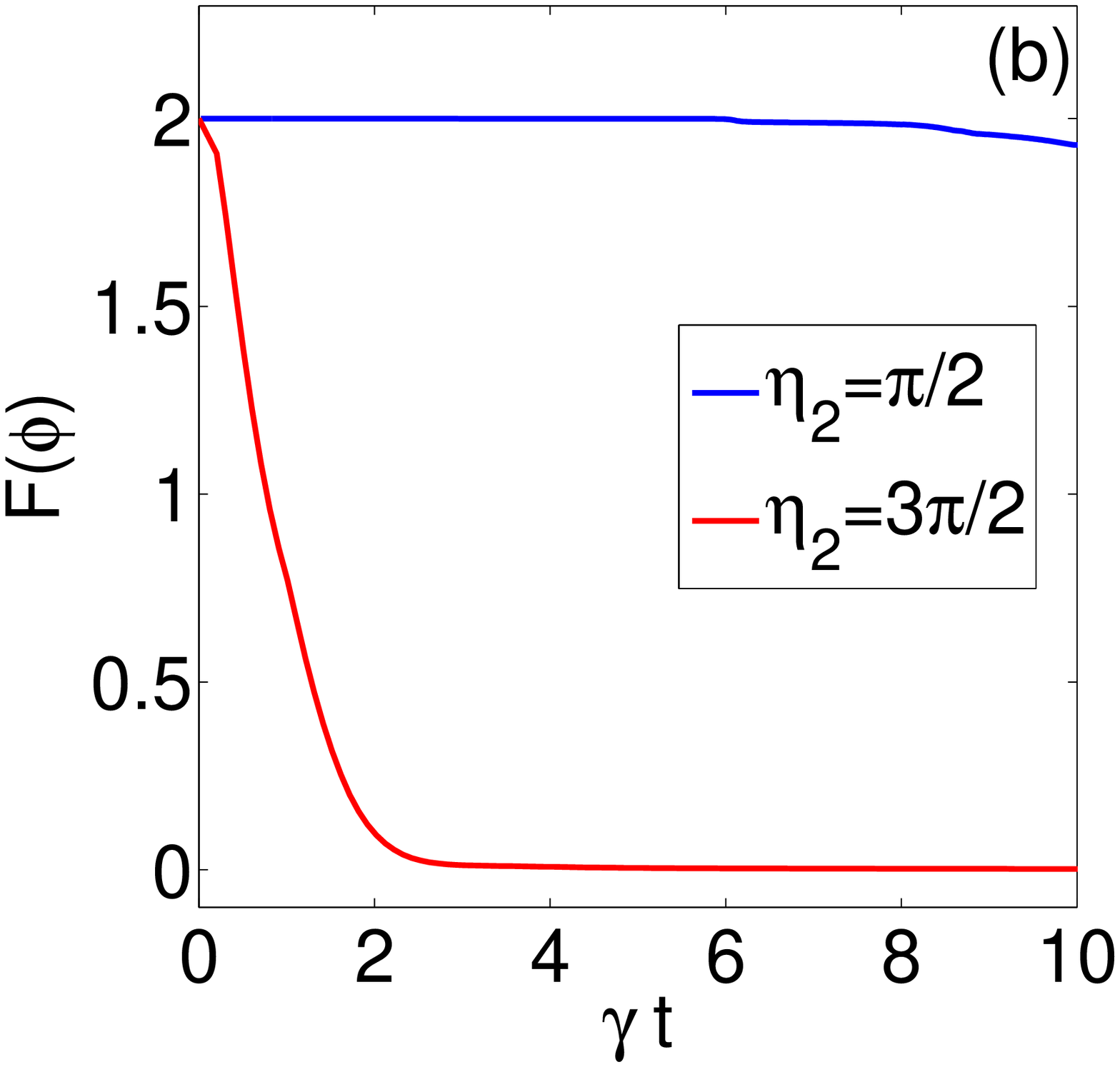}
  \end{minipage}
  \begin{minipage}[t]{0.5\linewidth}
    \includegraphics[width=1.7in,height=4.3cm]{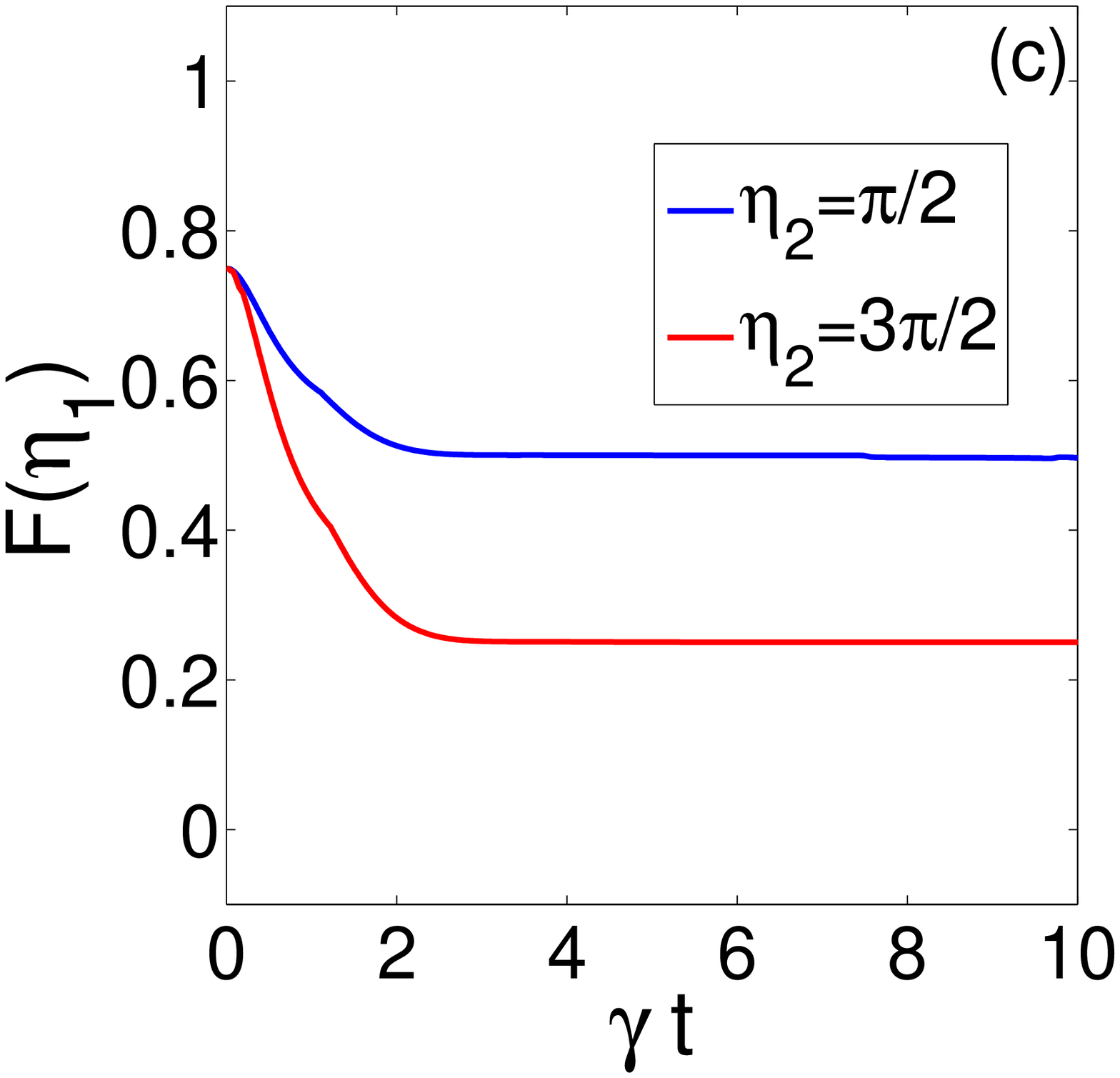}
  \end{minipage}%
  \begin{minipage}[t]{0.5\linewidth}
    \includegraphics[width=1.7in,height=4.3cm]{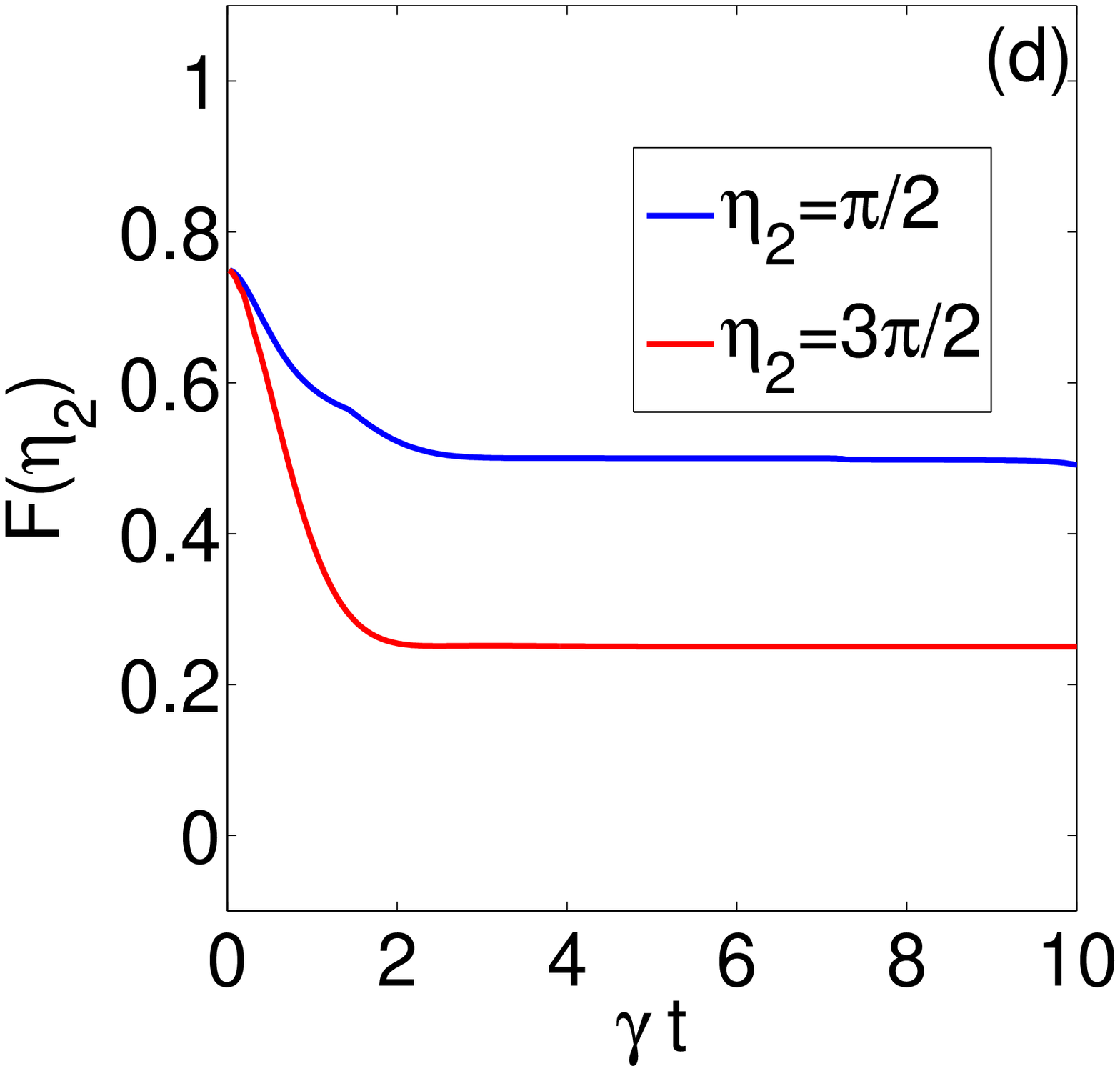}
  \end{minipage}
  \caption{\small Time evolution of QFI for the open V-type atom in the interference conditions, where $\gamma_{1}=\gamma_{2}\equiv\gamma$, $\lambda=2\gamma$, $\omega_{1}=\omega_{2}=\omega_{0}=91\gamma$. The QFIs are evaluated at $\theta=\phi=\pi/4$, $\eta_{1}=\pi/2$, $\eta_{2}=\pi/2$ (blue lines) or $\eta_{2}=3\pi/2$ (red lines).}
\end{figure}

\begin{figure}
  \begin{minipage}[t]{0.5\linewidth}
    \includegraphics[width=1.7in,height=4.3cm]{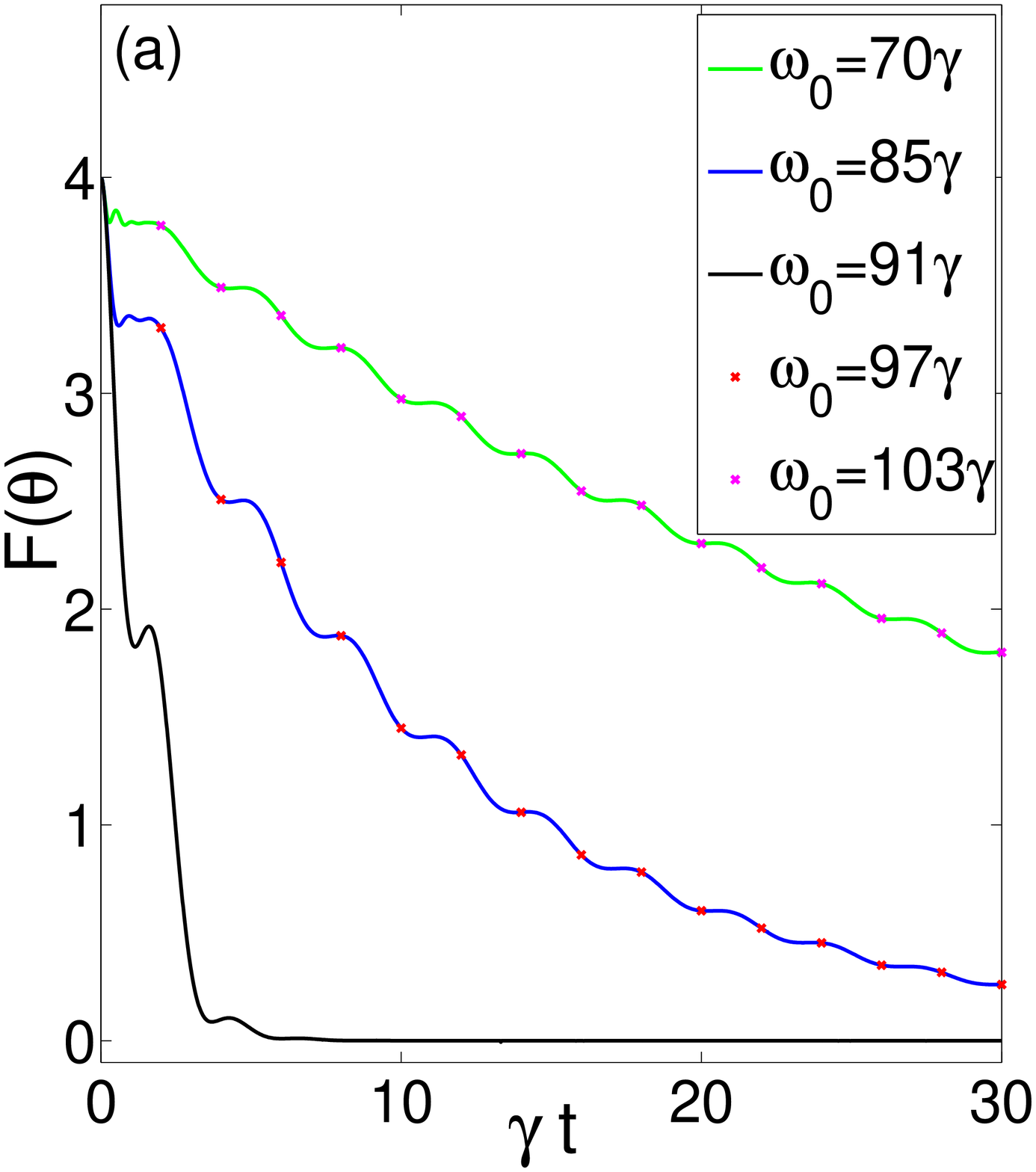}
  \end{minipage}%
  \begin{minipage}[t]{0.5\linewidth}
    \includegraphics[width=1.7in,height=4.3cm]{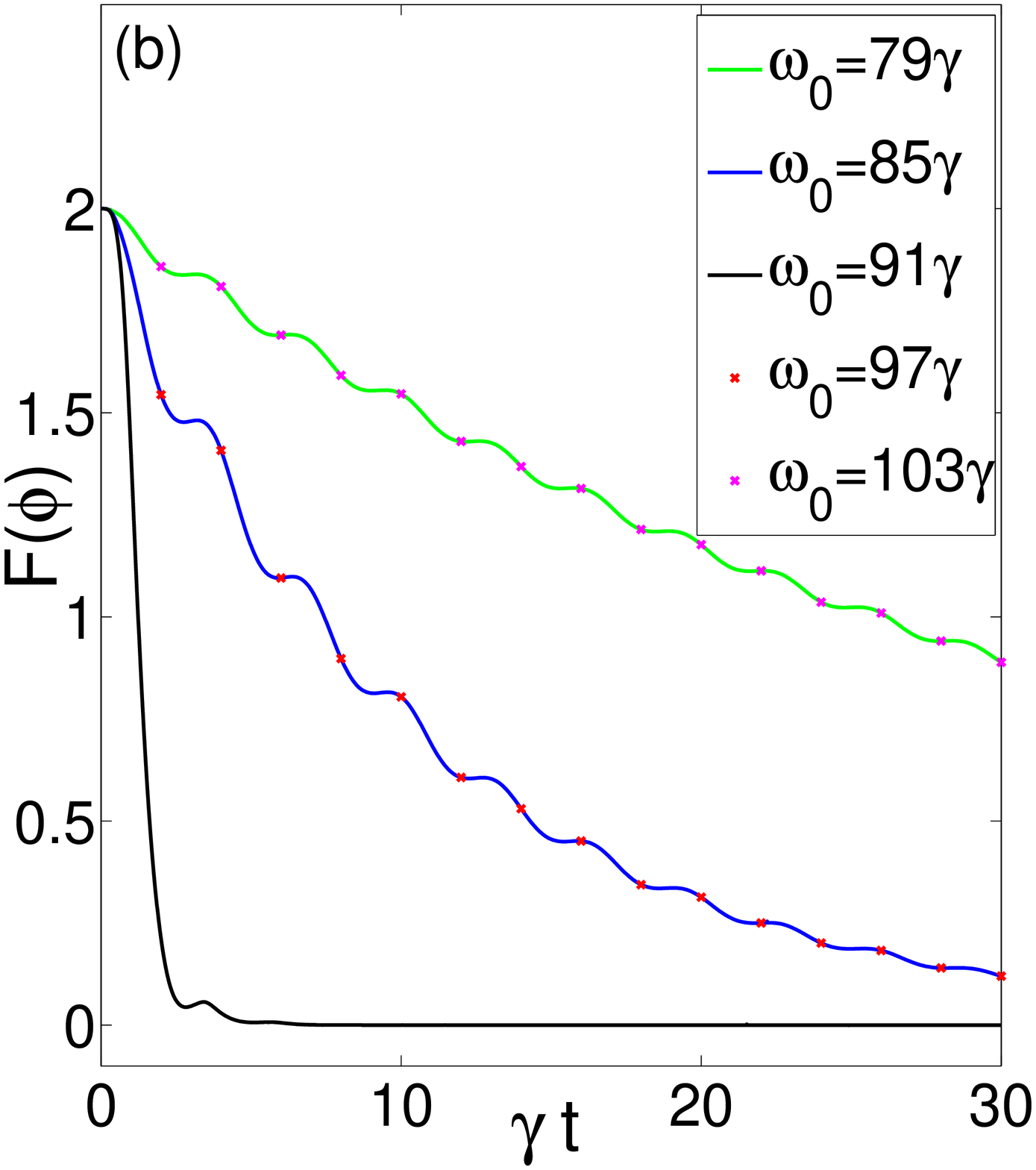}
  \end{minipage}
  \begin{minipage}[t]{0.5\linewidth}
    \includegraphics[width=1.7in,height=4.3cm]{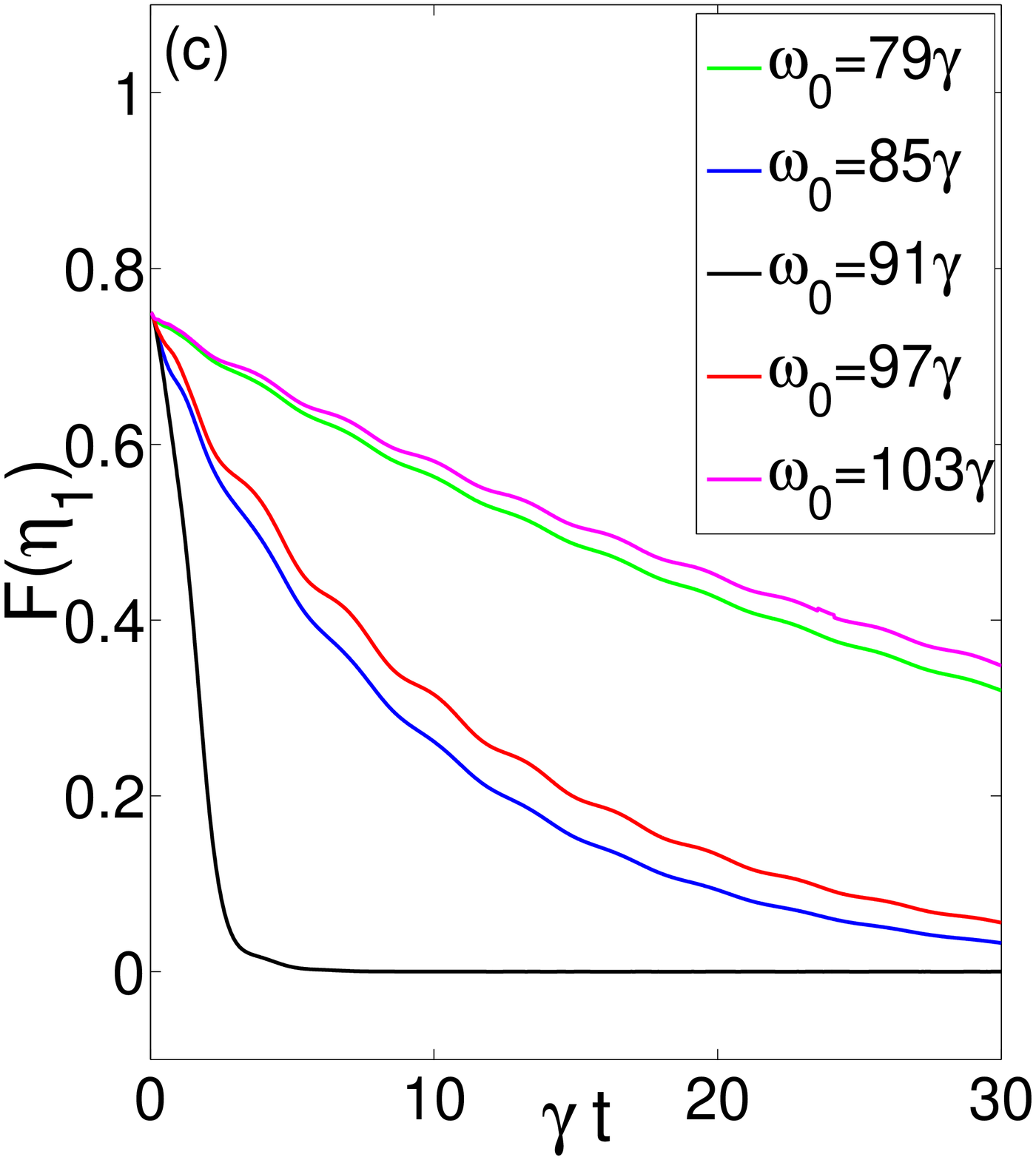}
  \end{minipage}%
  \begin{minipage}[t]{0.5\linewidth}
    \includegraphics[width=1.7in,height=4.3cm]{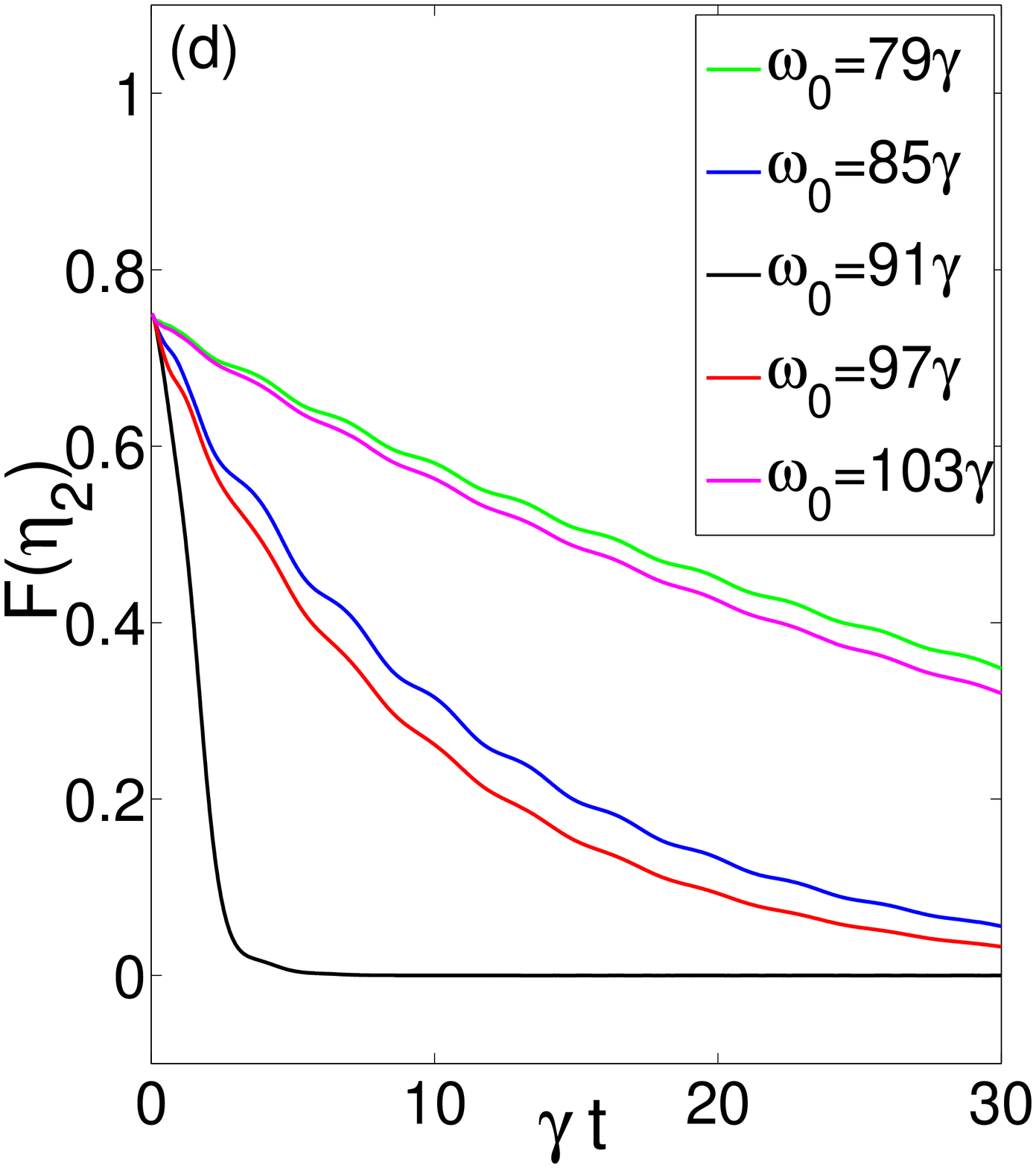}
  \end{minipage}
  \caption{\small Time evolution of QFI for the open V-type atom, where $\gamma_{1}=\gamma_{2}=\gamma$, $\lambda=2\gamma$, $\omega_{1}=90\gamma$, $\omega_{2}=92\gamma$. Five curves in each subgraph correspond to $\omega_{0}=79\gamma,85\gamma, 91\gamma, 97\gamma, 103\gamma$ respectively. The QFIs are evaluated at $\theta=\phi=\pi/4$, $\eta_{1}=\eta_{2}=\pi/2$.}
\end{figure}

\section{Evolution of quantum entanglement and quantum coherence}
As the last example of applications, we study the evolution of quantum entanglement and quantum coherence for the open V-type three-level atom. Entanglement and coherence describe the two different aspects of a quantum state--quantum correlation and purity. Both of them are the important resource in quantum information processing. We will find that the quantum interference has good protective roles to both the quantum entanglement and coherence.

We employ the notion of entanglement negativity as the description of quantum entanglement. For a bipartite system state $\rho_{\rm AB}$, entanglement negativity is defined as \cite{Peres,Horodecki}
\begin{equation}
   N(\rho_{\rm AB})=\sum_{k}|\eta_{k}^{T_{\rm A}(-)}|=\frac{\sum_{k}|\eta_{k}^{T_{\rm A}}|-1}{2},
\end{equation}
where $\eta_{k}^{T_{\rm A}(-)}$ and $\eta_{k}^{T_{\rm A}}$ are respectively the negative and all eigenvalues of the partial transpose of $\rho_{\rm AB}$ with respect to subsystem A.

Obviously, for studying the evolution of entanglement, the key step is to obtain the evolved state $\rho_{\rm AB}(t)$ of the open entangled system. For this end, we need to find the quantum dynamical map for the open quantum system. For the open V-type atom discussed above, we find from the Eqs. (4), (5a), (17a)-(17b) that the dynamical map $\varepsilon$ satisfies the roles
\begin{eqnarray}
 \nonumber \varepsilon (|0\rangle\langle0|)&=&|0\rangle\langle0|,\\
 \nonumber \varepsilon (|1\rangle\langle1|)&=&[1-|E(t)|^{2}-|H(t)|^{2}]|0\rangle\langle0|\\
 \nonumber&+&|E(t)|^{2}|1\rangle\langle1|+|H(t)|^{2}|2\rangle\langle2|\\
 \nonumber&+&E(t)H^{*}(t)|1\rangle\langle2|+E^{*}(t)H(t)|2\rangle\langle1|,\\
 \nonumber \varepsilon (|2\rangle\langle2|)&=&[1-|F(t)|^{2}-|G(t)|^{2}]|0\rangle\langle0|\\
 \nonumber&+&|F(t)|^{2}|1\rangle\langle1|+|G(t)|^{2}|2\rangle\langle2|\\
 \nonumber&+&F(t)G^{*}(t)|1\rangle\langle2|+F^{*}(t)G(t)|2\rangle\langle1|,\\
 \nonumber \varepsilon (|1\rangle\langle0|)&=&E(t)|1\rangle\langle0|+H(t)|2\rangle\langle0|,\\
 \nonumber \varepsilon (|2\rangle\langle0|)&=&F(t)|1\rangle\langle0|+G(t)|2\rangle\langle0|,\\
 \nonumber \varepsilon (|2\rangle\langle1|)&=&[-F(t)E^{*}(t)-G(t)H^{*}(t)]|0\rangle\langle0|\\
 \nonumber &+&F(t)E^{*}(t)|1\rangle\langle1|+G(t)H^{*}(t)|2\rangle\langle2|\\
 \nonumber &+&F(t)H^{*}(t)|1\rangle\langle2|+E^{*}(t)G(t)|2\rangle\langle1|.
\end{eqnarray}
Having this map in hand, we can calculate in principle the evolution of any quantum entangled state. Here we take the Werner-like state\cite{Werner},
\begin{equation}
    \rho_{\varepsilon}=\frac{(1-\varepsilon)}{9}\texttt{I}+\varepsilon|\Psi^{\rm AB}\rangle\langle\Psi^{\rm AB}|,
\end{equation}
as the exemplary example. Where $\texttt{I}$ denotes the 3-dimensional identity matrix, and  $|\Psi^{\rm AB}\rangle=\frac{1}{\sqrt{3}}(|00\rangle+|11\rangle+|22\rangle)$ is the maximally entangled state of two qutrits. The Werner-like state is separable for $0\leq \varepsilon \leq 1/4$, and entangled for $1/4<\varepsilon \leq 1$.

We discuss the problem in two cases: unilateral environment and bilateral environment. The former means that only the atom A is influenced by the noisy environment but atom B keeps noise-free, the later mean that both of the two atoms are influenced by noises. In Fig.8, we show the time evolution of the entanglement negativity of the Werner-like state for this two cases. It is shown that the entanglement negativity for both unilateral and bilateral environment have similar decaying behaviors. An interesting phenomenon is that the entanglement negativity reduces to zero in the case without quantum interference (solid lines), but to a nonzero asymptotic value in the case with quantum interference (dash lines). This result demonstrates that the quantum interference has good protective roles on quantum entanglement.

\begin{figure}
  \begin{minipage}[t]{0.5\linewidth}
    \includegraphics[width=1.7in,height=4.3cm]{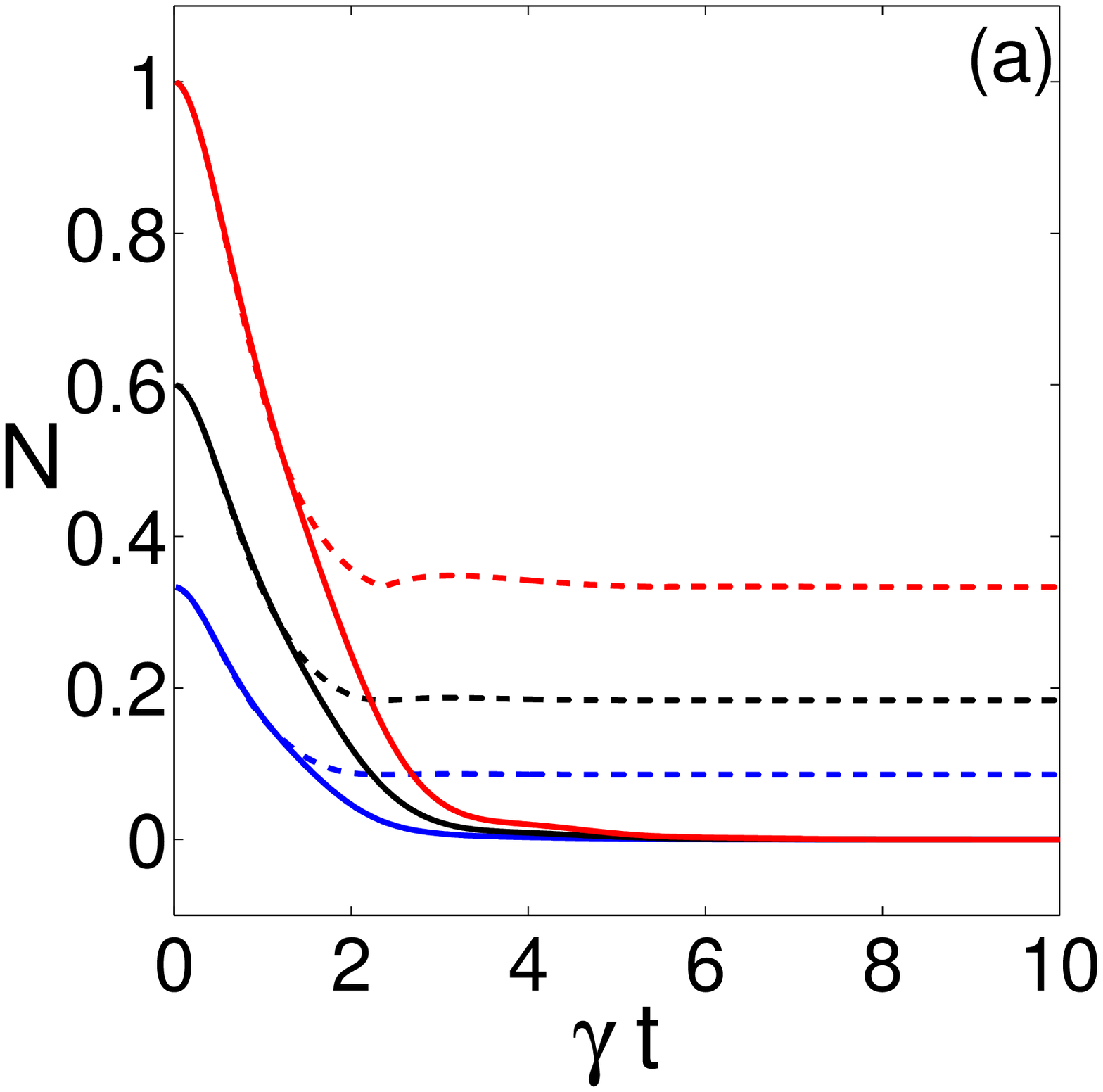}
  \end{minipage}%
  \begin{minipage}[t]{0.5\linewidth}
    \includegraphics[width=1.7in,height=4.3cm]{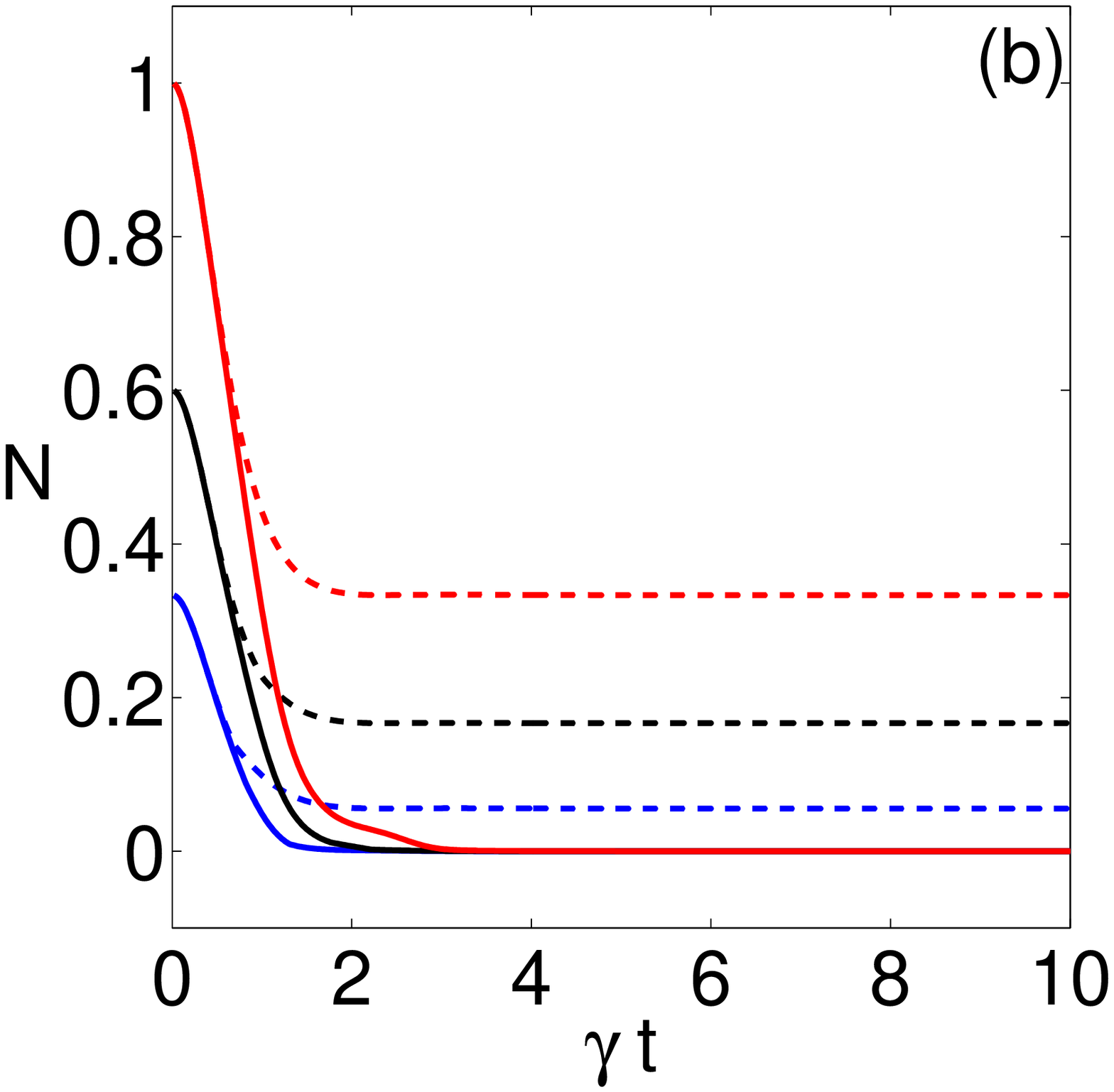}
  \end{minipage}
  \caption{Evolution of entanglement negativity versus dimensionless time $\gamma t$ for the open V-type atom with initial Werner-like state of Eq. (39). The parameters are chosen as $\gamma_{1}=\gamma_{2}=\gamma$, $\lambda=2\gamma$. (a) corresponds to unilateral environment and (b) corresponds to bilateral environment. The dash lines correspond to the case of quantum interference with $\omega_{1}=\omega_{2}=\omega_{0}=90\gamma$, and the solid lines correspond to the case without quantum interference with $\omega_{1}=90\gamma$, $\omega_{2}=92\gamma$, $\omega_{0}=91\gamma$. The red, black and blue lines correspond to $\varepsilon=1,0.7,0.5$ respectively.}
\end{figure}

Another important quantity that describes quantum states is the quantum coherence. It is a very important notion in quantum physics, but a rigorous quantification of it has been lacked. Up to very recently, Baumgratz, Cramer and Plenio \cite{Baumgratz2014} established a rigorous framework for the quantification of coherence from the point of resource theory. Two typical measures, i.e., the $l_1$ norm of coherence and the relative entropy of coherence (REC) in the framework, were presented. In a reference basis $\left\{
{\left| i \right\rangle } \right\}_{i = 1, \ldots ,d}$ of a
$d$-dimensional quantum system, the $l_1$ norm of coherence is simply defined as the sum of the absolute value of all the off-diagonal elements of the system density matrix,
\begin{equation}
    C_{l_{1}}(\rho)=\sum_{\substack{i,j\\i\neq j}}|\rho_{i,j}|.
\end{equation}
Obviously, this is a very simple and intuitive definition. The
REC
is defined as
\begin{equation}
C_{\rm rel} \left( \rho  \right) = S\left( {\rho
_{\rm{diag}} } \right) - S\left( \rho  \right),
\end{equation}
where $S$ is the von Neumann entropy function and $\rho_{\rm diag}$ denotes the state obtained from $\rho$ by deleting all off-diagonal elements.
Though both the two measures of quantum coherence satisfy the requirements of the resource theory, are they really compatible or equivalent?

Fig.9 gives the time evolution of the two kinds of coherence measures for the Werner-like state Eq.(39) of the V-type atomic system. It is shown that the two measures of coherence are not always compatible. For example, in the beginning stage of Fig.9(b), the blue dash line is increased but the red dash line is decreased; the blue solid line increases firstly and then decreases, but the red solid line drops directly. Fig.9(d) also reveals distinct differences for the two kinds of coherence measures.

Another interesting result revealed by Fig.9 is that though the coherence reduces quickly in the case of having no quantum interference, it has large asymptotic value in the case of quantum interference. The asymptotic value is even much larger than its initial value. This is to say that quantum interference can enhance and protect effectively the coherence of quantum states.

\begin{figure}
  \begin{minipage}[t]{0.5\linewidth}
    \includegraphics[width=1.7in,height=4.3cm]{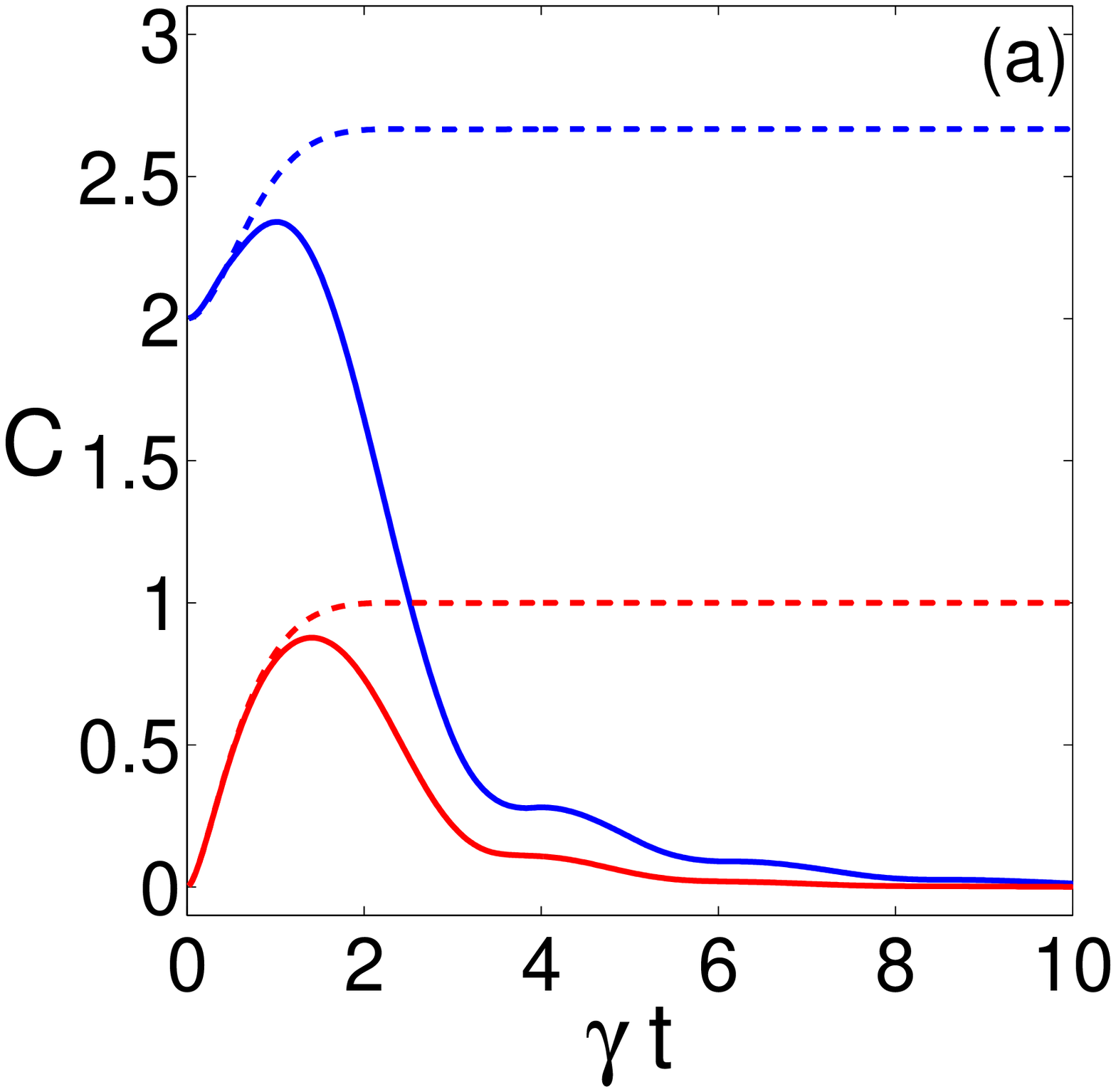}
  \end{minipage}%
  \begin{minipage}[t]{0.5\linewidth}
    \includegraphics[width=1.7in,height=4.3cm]{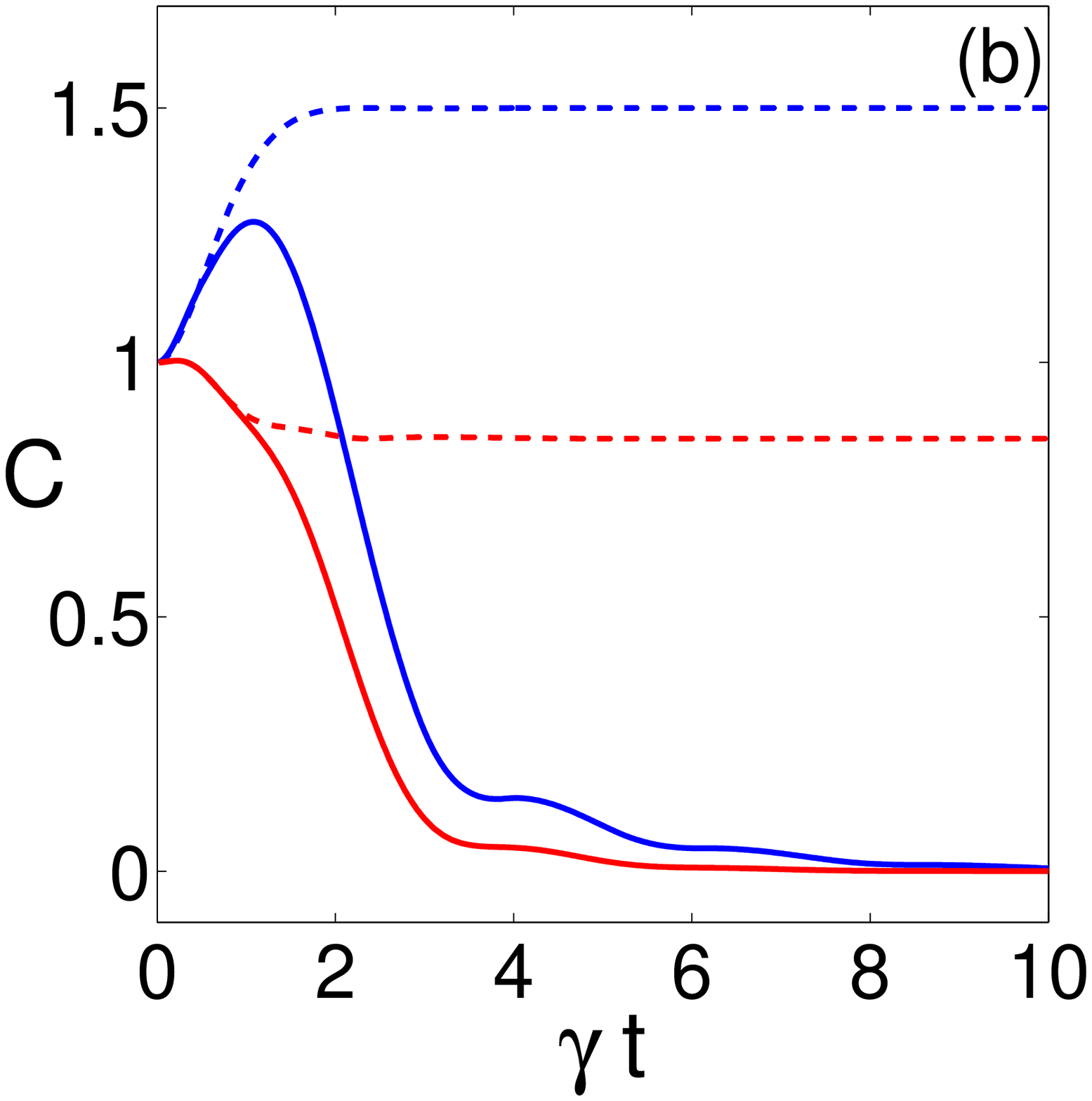}
  \end{minipage}
  \begin{minipage}[t]{0.5\linewidth}
    \includegraphics[width=1.7in,height=4.3cm]{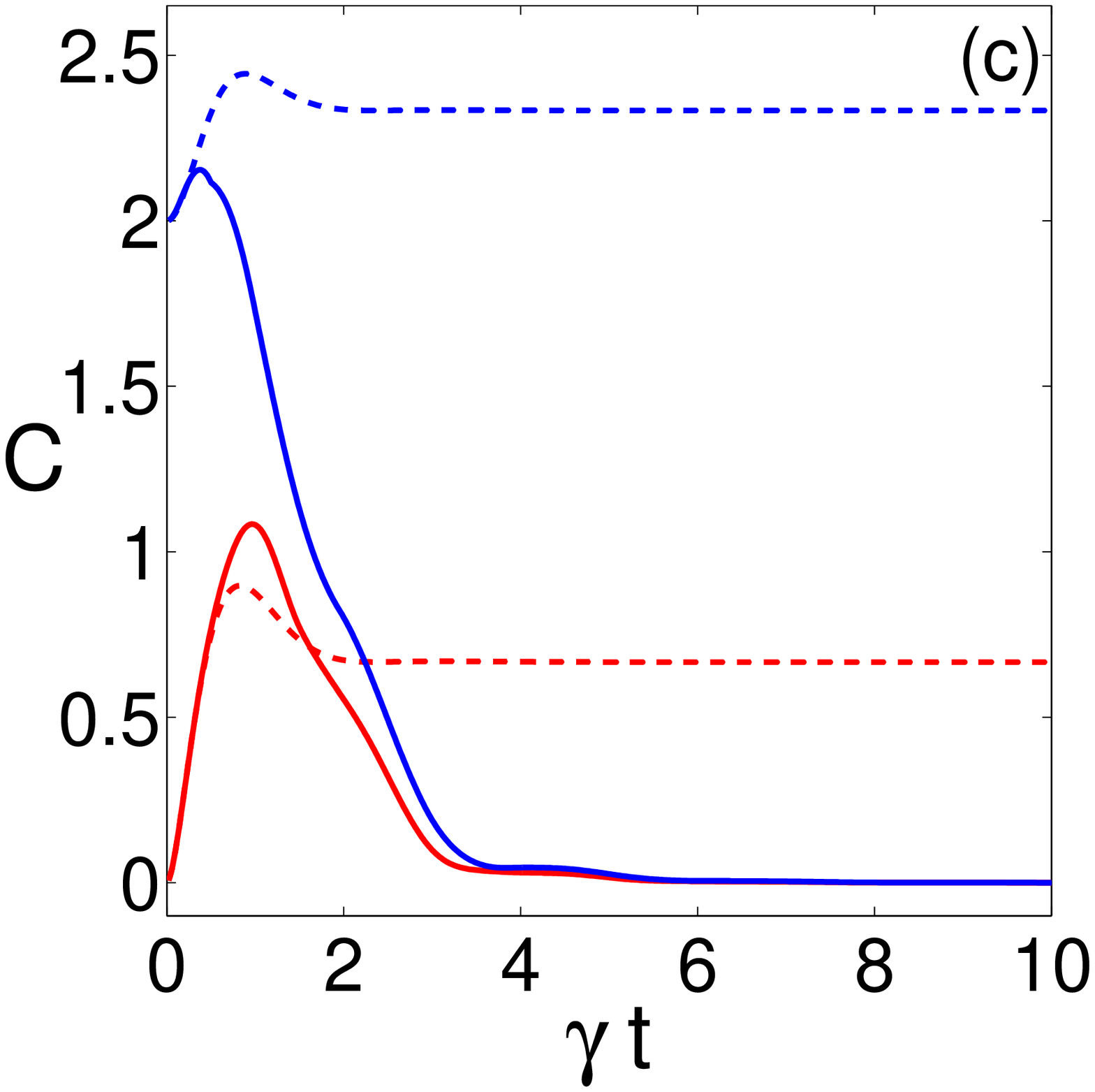}
  \end{minipage}%
  \begin{minipage}[t]{0.5\linewidth}
    \includegraphics[width=1.7in,height=4.3cm]{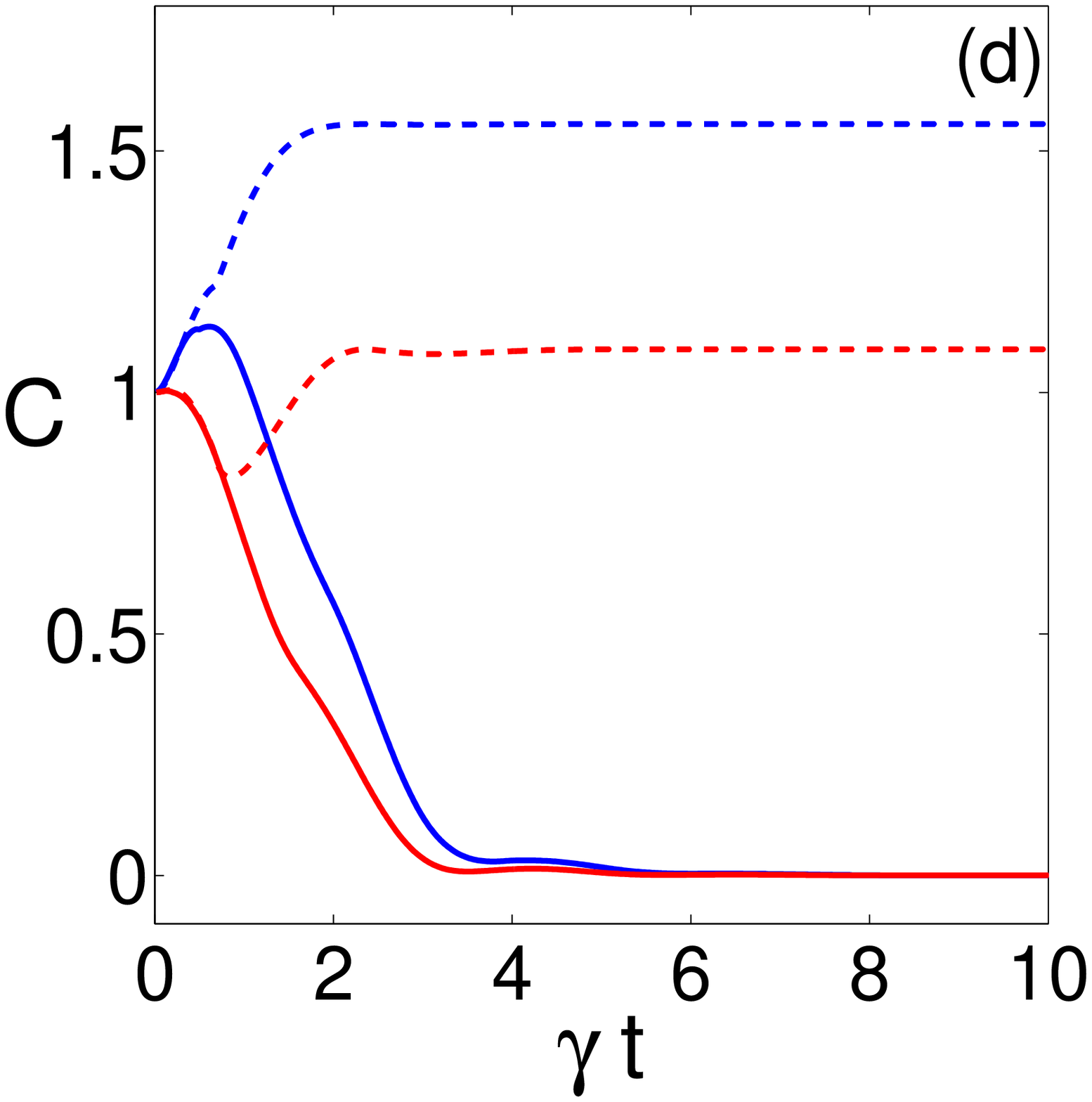}
  \end{minipage}
  \caption{\small Time evolution of quantum coherence for the open V-type atom with initial Werner-like state, where $\gamma_{1}=\gamma_{2}=\gamma$, $\lambda=2\gamma$. (a) unilateral environment with $\varepsilon=1$; (b) unilateral environment with $\varepsilon=0.5$; (c) bilateral environment with $\varepsilon=1$; (d) bilateral environment with $\varepsilon=0.5$; The blue lines correspond to $C_{l_{1}}$ and red lines correspond to REC. Dash lines correspond to the case of quantum interference with $\omega_{1}=\omega_{2}=\omega_{0}=90\gamma$, and solid lines to the case without quantum interference with $\omega_{1}=90\gamma$, $\omega_{2}=92\gamma$, $\omega_{0}=91\gamma$.}
\end{figure}

\section{conclusions}
In conclusion, we have presented the exactly analytical solutions for the dynamics of the dissipative three-level V-type and $\Lambda$-type atomic systems in the vacuum Lorentzian environments. We have then discussed the phenomenon of quantum interference for the two kinds of dynamical models.  Especially, the quantum interference conditions between the transitions of V-type atomic system have been derived. Finally, by taking the open V-type three-level atom as the exemplum, we have studied the dynamical evolution of quantum Fisher information, quantum entanglement and quantum coherence, especially highlighting the roles of quantum interference.

Starting from the property of the asymptotic populations, we have derived the necessary conditions of the quantum interference between the two decaying transitions for open V-type atom. They are completely consistent with the interference conditions of classical light. These interference phenomena have been examined further by numerical simulations, and especially the destructive and constructive interferences have been observed. We have also theoretically demonstrated that the similar phenomenon of quantum interference does not exist in the dissipative $\Lambda$-type atomic system. We believe that these results are important in the theory of quantum optics.

Quantum coherence, quantum entanglement and quantum Fisher information are the very important notions in quantum mechanics. They are important physical resources in quantum information processing. We have demonstrated that the quantum interference for the dissipative V-type atomic system can protect effectively the quantum entanglement and quantum coherence, though it is not always valid for the protection of quantum Fisher information. We have also demonstrated that the two typical measures of quantum coherence, presented recently by Baumgratz, Cramer and Plenio \cite{Baumgratz2014} from the point of resource theory, are incompatible in principle. This result implying that the study to the essence of quantum coherence has still a long way to go.

We have still made some other results. For example, the memory effect is beneficial to slower the decaying of quantum Fisher information, and make the decaying curves to take on some oscillations. Especially, when the center frequency of the Lorentzian environment is located at the middle between the two transition frequencies of the V-type atom, the decaying of the quantum Fisher information is the fastest. Deviating from this middle frequency to the two sides, the decaying becomes slower and slower.

It is worthwhile to point out that the method for solving the dynamics can in principle be generalized to the case of more than three-level atomic systems, as long as the condition of one single excitation is assumed. For example in the case of open multilevel $V$-type atom, following the process for dealing with the open three-level V-type atom presented in Sec.II, one will obtain a set of coupled integro-differential equations with respect to coefficients $c_{i}(t)$, like Eqs. (8)-(9). Their Laplace transformation, under the Lorentzian spectrum, can be written in the form of $c_{i}(p)=A(p)/B(p)$ [like Eqs. (13)-(14)], where both $A(p)$ and $B(p)$ are polynomials of $p$ with the highest power of $B(p)$ larger than that of $A(p)$. The inverse Laplace transformation of $C_{i}(p)$ is thus feasible, following the method in the text. Similar trick may also be used in the open N-type or M-type atomic systems under the condition of one single excitation.

\begin{acknowledgments}
This work is supported by the National Natural
Science Foundation of China (Grant No. 11275064), the Specialized Research Fund for the Doctoral Program of Higher Education (Grant No. 20124306110003), and the Construct Program of the
National Key Discipline.
\end{acknowledgments}

\renewcommand{\theequation}{A.\arabic{equation}}
\setcounter{equation}{0}
\section*{Appendix A: The inverse Laplace transform of Eqs.(13)-(14) for the degenerate cases}

If the polynomials $p^{3}+h_{1}p^{2}+h_{2}p+h_{3}=0$ has a two-fold root $b_{1}$ and a single root $b_{3}$, then the decomposition of Eqs.(16) and (17) becomes
\begin{subnumcases}{}
  c_{1}(p)=\frac{\hat{D}_{2}}{(p-b_{1})^{2}}+\frac{\hat{D}_{1}}{p-b_{1}}+\frac{\hat{D}_{3}}{p-b_{3}},\\
  c_{2}(p)=\frac{\hat{D}'_{2}}{(p-b_{1})^{2}}+\frac{\hat{D}'_{1}}{p-b_{1}}+\frac{\hat{D}'_{3}}{p-b_{3}},
\end{subnumcases}
where
\begin{subnumcases}{}
  \hat{D}_{i}=\hat{E}_{i}c_{1}(0)+\hat{F}_{i}c_{2}(0),\\
  \hat{D}'_{i}=\hat{G}_{i}c_{2}(0)+\hat{H}_{i}c_{1}(0),
\end{subnumcases}
with
\begin{eqnarray}
 \nonumber \hat{E}_{1}&=&\frac{b_{1}^{2}-(2b_{1}+M+{\rm i}\omega_{2})b_{3}-{\rm i}\omega_{2}M-B_{22}}{(b_{1}-b_{3})^{2}},\\
 \nonumber \hat{F}_{1}&=&\frac{B_{12}}{(b_{1}-b_{3})^{2}},\\
 \nonumber \hat{E}_{2}&=&\frac{(b_{1}+{\rm i}\omega_{2})(b_{1}+M)+B_{22}}{b_{1}-b_{3}},\\
 \nonumber \hat{F}_{2}&=&-\frac{B_{12}}{b_{1}-b_{3}},\\
 \nonumber \hat{E}_{3}&=&\frac{(b_{3}+{\rm i}\omega_{2})(b_{3}+M)+B_{22}}{(b_{1}-b_{3})^{2}},\\
 \nonumber \hat{F}_{3}&=&-\frac{B_{12}}{(b_{1}-b_{3})^{2}},
\end{eqnarray}
and
\begin{eqnarray}
 \nonumber \hat{G}_{1}&=&\frac{b_{1}^{2}-(2b_{1}+M+{\rm i}\omega_{1})b_{3}-{\rm i}\omega_{1}M-B_{11}}{(b_{1}-b_{3})^{2}},\\
 \nonumber \hat{H}_{1}&=&\frac{B_{21}}{(b_{1}-b_{3})^{2}},\\
 \nonumber \hat{G}_{2}&=&\frac{(b_{1}+{\rm i}\omega_{1})(b_{1}+M)+B_{11}}{b_{1}-b_{3}},\\
 \nonumber \hat{H}_{2}&=&-\frac{B_{21}}{b_{1}-b_{3}},\\
 \nonumber \hat{G}_{3}&=&\frac{(b_{3}+{\rm i}\omega_{1})(b_{3}+M)+B_{11}}{(b_{1}-b_{3})^{2}},\\
 \nonumber \hat{H}_{3}&=&-\frac{B_{21}}{(b_{1}-b_{3})^{2}}.
\end{eqnarray}
Finally the inverse Laplace transform of equations (A.1a) and (A.1b) gives
\begin{subnumcases}{}
  c_{1}(t)=\hat{E}(t)c_{1}(0)+\hat{F}(t)c_{2}(0),\\
  c_{2}(t)=\hat{G}(t)c_{2}(0)+\hat{H}(t)c_{1}(0),
\end{subnumcases}
with
\begin{eqnarray}
 \nonumber \hat{E}(t)&=&(\hat{E}_{1}+\hat{E}_{2}t)e^{b_{1}t}+\hat{E}_{3}e^{b_{3}t},\\
 \nonumber \hat{F}(t)&=&(\hat{F}_{1}+\hat{F}_{2}t)e^{b_{1}t}+\hat{F}_{3}e^{b_{3}t},\\
 \nonumber \hat{G}(t)&=&(\hat{G}_{1}+\hat{G}_{2}t)e^{b_{1}t}+\hat{G}_{3}e^{b_{3}t},\\
 \nonumber \hat{H}(t)&=&(\hat{H}_{1}+\hat{H}_{2}t)e^{b_{1}t}+\hat{H}_{3}e^{b_{3}t}.
\end{eqnarray}

If the polynomials $p^{3}+h_{1}p^{2}+h_{2}p+h_{3}=0$ has only one three-fold root $b$, then Eqs.(16) and (17) become
\begin{subnumcases}{}
  c_{1}(p)=\frac{\check{D}_{3}}{(p-b)^{3}}+\frac{\check{D}_{2}}{(p-b)^{2}}+\frac{\check{D}_{1}}{p-b},\\
  c_{2}(p)=\frac{\check{D}'_{3}}{(p-b)^{3}}+\frac{\check{D}'_{2}}{(p-b)^{2}}+\frac{\check{D}'_{1}}{p-b},
\end{subnumcases}
where
\begin{eqnarray}
 \nonumber \check{D}_{3}&=&[(b+{\rm i}\omega_{2})(b+M)+B_{22}]c_{1}(0)-B_{12}c_{2}(0),\\
 \nonumber \check{D}_{2}&=&(2b+M+{\rm i}\omega_{2})c_{1}(0),\\
 \nonumber \check{D}_{1}&=&2c_{1}(0),\\
 \nonumber \check{D}'_{3}&=&[(b+{\rm i}\omega_{1})(b+M)+B_{11}]c_{2}(0)-B_{21}c_{1}(0),\\
 \nonumber \check{D}'_{2}&=&(2b+M+{\rm i}\omega_{1})c_{2}(0),\\
 \nonumber \check{D}'_{1}&=&2c_{2}(0).
\end{eqnarray}
The inverse Laplace transform of equations (A.4a) and (A.4b) gives
\begin{subnumcases}{}
  c_{1}(t)=\check{E}(t)c_{1}(0)+\check{F}(t)c_{2}(0),\\
  c_{2}(t)=\check{G}(t)c_{2}(0)+\check{H}(t)c_{1}(0),
\end{subnumcases}
with
$\check{E}(t)=\{\frac{1}{2}[(b+{\rm i}\omega_{2})(b+M)+B_{22}]t^{2}+(2b+M+{\rm i}\omega_{2})t+2\}e^{bt}$, $\check{F}(t)=-\frac{1}{2}B_{12}t^{2}e^{bt}$, $\check{G}(t)=\{\frac{1}{2}[(b+{\rm i}\omega_{1})(b+M)+B_{11}]t^{2}+(2b+M+{\rm i}\omega_{1})t+2\}e^{bt}$, $\check{H}(t)=-\frac{1}{2}B_{21}t^{2}e^{bt}$.

\renewcommand{\theequation}{B.\arabic{equation}}
\setcounter{equation}{0}
\section*{Appendix B: Proof of no real root of Eqs.(36a)-(36b)}
If $\lambda\neq 0$, then $\chi\neq 0$. Multiplying Eq.(36a) by $\chi/2$ and then subtracting Eq.(36b), one has
\begin{equation}
  (\delta_{1}+\delta_{2})\chi^{2}-[2\lambda^{2}-2\delta_{1}\delta_{2}+\frac{\lambda}{2}(\gamma_{1}+\gamma_{2})]\chi-\lambda(\gamma_{1}\delta_{2}+\gamma_{2}\delta_{1})=0
\end{equation}
If $\delta_{1}+\delta_{2}\neq 0$, multiplying Eq.(36a) by $(\delta_{1}+\delta_{2})/2$ and then subtracting Eq.(B.1), one gets
\begin{equation}
 \chi=\frac{\lambda\gamma_{1}(3\delta_{2}-\delta_{1})+\lambda\gamma_{2}(3\delta_{1}-\delta_{2})}{8\lambda^{2}+2(\delta_{1}-\delta_{2})^{2}+2\lambda(\gamma_{1}+\gamma_{2})}.
\end{equation}
Plugging it into Eq.(36a), we can obtain the quadratic equation with respect to $\omega_{0}$,
\begin{equation}
  a\omega_{0}^{2}+b\omega_{0}+c=0
\end{equation}
where
$  a=4\lambda^{2}(u+v)^{2}+4\lambda n(u+v)$,
$ b=-8\lambda^{2}(u+v)(\omega_{1}u+\omega_{2}v)-2\lambda n[(3\omega_{1}+\omega_{2})u+(3\omega_{2}+\omega_{1})v]$ and
$c=4\lambda^{2}(\omega_{1}u+\omega_{2}v)^{2}+2\lambda n(\omega_{1}+\omega_{2})(\omega_{1}u+\omega_{2}v)+\frac{1}{2}\lambda n^{2}(u+v)$, with $u=\gamma_{1}-3\gamma_{2}$, $v=\gamma_{2}-3\gamma_{1}$ and $n=2(\omega_{1}-\omega_{2})^{2}+8\lambda^{2}+2\lambda (\gamma_{1}+\gamma_{2})$. The discriminant of Eq.(B.3) is
\begin{eqnarray}
  \nonumber\Delta&=&b^{2}-4ac\\
 \nonumber &=&-256\lambda^{2}n^{2}[\gamma_{1}\gamma_{2}(\omega_{1}-\omega_{2})^{2}+\lambda^{2}(\gamma_{1}+\gamma_{2})^{2}]<0,
\end{eqnarray}
which implies that $\omega_{0}$ is a complex number.

If $\delta_{1}+\delta_{2}= 0$, then Eqs.(36a) and (B.1) reduce respectively to
\begin{equation}
   4\chi^{2}-\lambda (\gamma_{1}+\gamma_{2})=0,
\end{equation}
\begin{equation}
   [4\lambda^{2}+4\delta_{1}^{2}+\lambda (\gamma_{1}+\gamma_{2})]\chi-\lambda\delta_{1}(\gamma_{1}-\gamma_{2})=0,
\end{equation}
which lead to the quadratic equation with respect to $\delta_{1}$
\begin{eqnarray}
 \nonumber \pm\sqrt{(\gamma_{1}+\gamma_{2})\lambda}\delta_{1}^{2}-&\lambda& (\gamma_{1}-\gamma_{2})\delta_{1}\pm\sqrt{(\gamma_{1}+\gamma_{2})\lambda^{5}}\\
  &\pm &\frac{1}{4}[(\gamma_{1}+\gamma_{2})\lambda]^{3/2}=0.
\end{eqnarray}
This equation also leads to $\delta_{1}$ having only complex roots.

\end{document}